\documentclass{aa}

\usepackage{graphicx}
\usepackage{amssymb,amsmath}
\usepackage{threeparttable}
\usepackage{txfonts}
\usepackage{float}
\usepackage{subfigure}
\usepackage{tabularx}
\usepackage{epsfig}

\def\msun {$\rm{M_{\odot}}$}
\def\kms {$\rm{km~s^{-1}}$}

\def\halfa{H$\alpha$}

\def\halfa{H$\alpha$~}

\begin{document}
 
\title{Supersonic turbulence in Giant HII Regions: clues from 30~Doradus
\thanks{Partly based on observations made with ESO Telescopes at the La Silla Paranal Observatory under programmes 072.C-0348; 182.D-0222; and 60.A-9700(G), and programme ID 076.C-0888, processed and released by the ESO VOS/ADP group.}
}

\author{ J.~Melnick\inst{1,2}, G.~Tenorio-Tagle\inst{3}, E.~Telles\inst{2}}
\institute{European Southern Observatory, Av. Alonso de Cordova 3107, Santiago, Chile,
\and
Observatorio Nacional, Rua Jos\'e Cristino 77, 20921-400 Rio de Janeiro, Brasil,
\and
Instituto Nacional de Astrof{\'i}sica {\'O}ptica y Electr{\'o}nica, L.E.Erro 1, Sta Mar{\'i}a Tonanzintla, 72840 San Andr{\'e}s Cholula, Pue. M{\'e}xico,
}

\offprints{Jorge Melnick \email{jmelnick@eso.org}}
\date{}

\authorrunning{Melnick et al.}

\titlerunning{Turbulence in Giant HII Regions}

\abstract{The tight correlation between turbulence and luminosity in Giant HII Regions is not well understood. While the luminosity is due to the UV radiation from the massive stars in the ionizing clusters, it is not clear what powers the turbulence. Observations of the two prototypical Giant HII Regions in the local Universe, 30~Doradus and NGC604, show that part of the kinetic energy of the nebular gas comes from the combined stellar winds of the most massive stars - the cluster winds, but not all.  We present a study of the kinematics of  30~Doradus based on archival VLT FLAMES/GIRAFFE data and new high resolution observations with HARPS. We find that the nebular structure and kinematics are shaped by a hot cluster wind and not by the stellar winds of individual stars. The cluster wind powers most of the turbulence of the nebular gas, with a small but significant contribution from the combined gravitational potential of stars and gas.  We estimate the total mass of 30~Doradus and we argue that the region does not contain significant amounts of neutral (HI) gas, and that the giant molecular cloud 30Dor-10 that is close to the center of the nebula in projection is in fact an inflating cloud tens of parsecs away from R136, the core of the ionizing cluster. We rule out a Kolmogorov-like turbulent kinetic energy cascade as the source of supersonic turbulence in Giant HII Regions.}
 

\maketitle

\section{Introduction}
\label{intro}

Besides the obvious difference in size, the fundamental distinction between Giant HII Regions (GHRs) and their smaller (mostly) Galactic counterparts (e.g. Orion) lies in the velocity width of their integrated emission-line profiles. Whereas small HII regions have thermal velocity widths, the integrated widths of GHRs are highly supersonic. 

Yet, despite decades of research and discussions, there is still no agreement as to the origin of the supersonic turbulence in GHRs. As so many other human endeavors of our days, opinions are divided into two seemingly irreconcilable camps: the "gravity" camp, championed by the authors of this paper in various combinations many with other collaborators (see e.g. \citealt{arenas} for a recent discussion and historical perspective); and the "stellar-winds" camp led by Y.H. Chu and R. Kennicutt \citep{chuken}.

\begin{figure}[h!]
\vspace{3.2cm}
\includegraphics[height=7cm,trim= 1.cm 0 8cm 10cm]{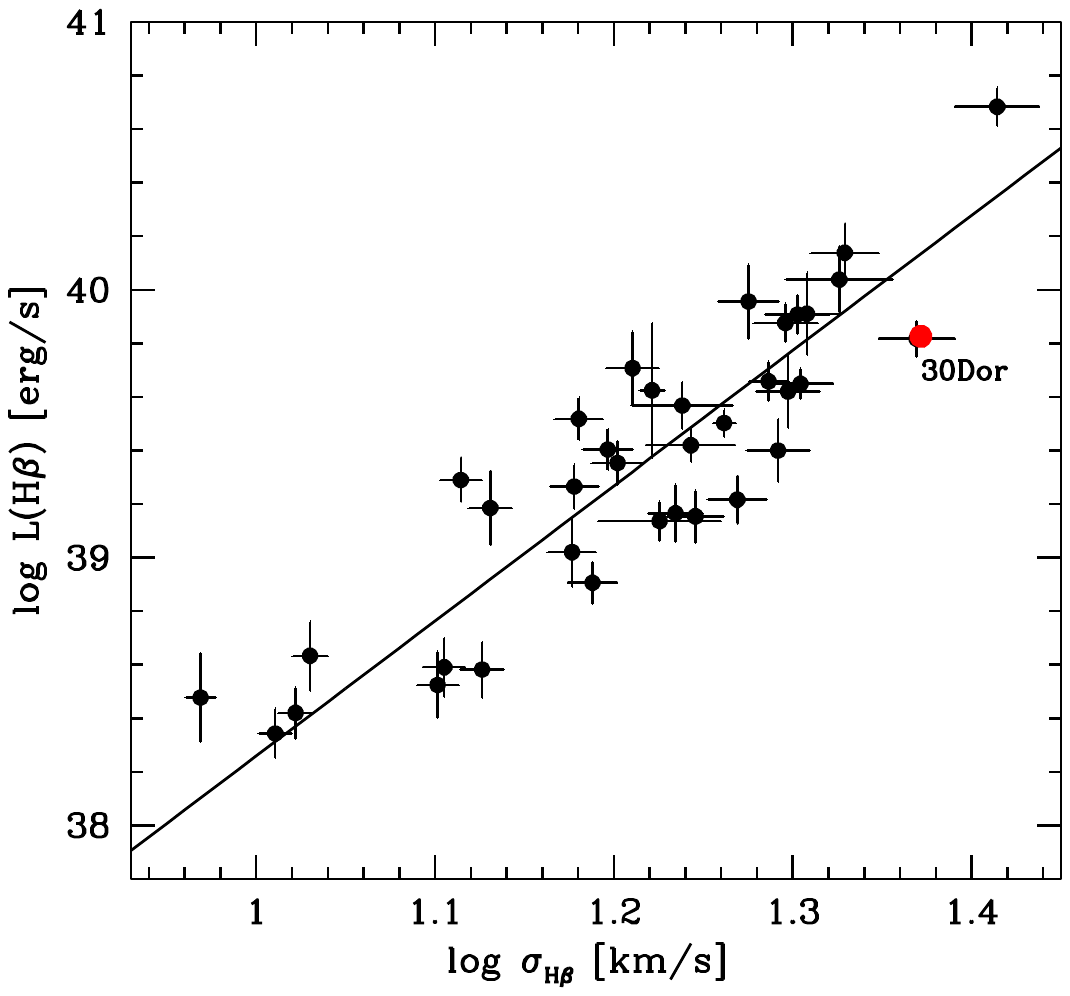} 
\vspace{-2.9cm}
\caption {The L-$\sigma$ relation for Giant HII Regions adapted from \cite{arenas}.}
\label{lsigma}
\end{figure}

The defenders of gravity argue that the surprisingly tight correlation between the global Balmer line-profile widths ($\sigma$;  H$\alpha$ or H$\beta$) and the integrated flux of these lines - the so-called L-$\sigma$ relation - (shown in Figure~\ref{lsigma}) is the unequivocal telltale of gravity, although the slope of the correlation ($L\propto\sigma^5$) cannot be deducted from first principles. The fact that HII region-like galaxies, or HII Galaxies also define a tight $L-\sigma$ relation of similar slope, is also an argument for the gravity camp.

The stellar winds camp, on the other side, base their arguments on  the morphology of GHRs, which is dominated by prominent arcs and filaments, and on high-resolution spectroscopy that shows that the emission lines associated with these morphological features are double or multiple as expected for expanding bubbles of ionized gas propelled by stellar winds. 

\cite{chuken} showed that when the ragged shapes of the spatially resolved profiles are added-up, the resulting integrated profiles have a smooth supersonic Gaussian core with broad unresolved wings matching surprisingly well observations of unresolved nebulae.

The idea that winds from individual stars can generate the observed bubbles has been disputed by \cite{silich} who argue that the close proximity of the massive stars within the ionizing clusters of GHRs implies that their winds must necessarily merge into a hot {\em cluster} wind that percolates the pre-existing interstellar medium generating thus the hierarchy of expanding shells that we see today.  Although the difference between stellar-winds and a cluster-wind is fundamental, it does not change the original argument of \cite{chuken} to explain the multiple shell structure of GHRs. 

Much of the work of the "stellar-wind" camp is based on observations of the two nearest and therefore best studied GHRs in the Local Group: 30~Doradus in the LMC and NGC604 in M33, but specially of 30~Doradus, which has become a sort of {\it Rosetta Stone} for deciphering the astrophysics of giant HII regions and their ionizing stars. In fact, the ionizing cluster of 30~Doradus (sometimes referred to as NGC2070; others as the R136 cluster), contains the largest concentration of resolved massive stars in the local Universe, including some of the few supermassive stars known (see \citealt{castro} for a recent review).

\cite{chuken} argued that about 50\% of the kinetic energy of the nebular gas in 30Dor is in the expanding shells and 50\% in some unspecified form of general turbulence. The mass of stars+gas in the nebula, ($M\sim5\times10^5$\msun)\ they argued, is not enough to generate the observed turbulence through the gravitational potential. 

The aim of the present paper is to use new public FLAMES/GIRAFFE data to investigate in (renewed) detail the nature of the supersonic turbulence of the nebular gas in 30~Doradus. 
 
\section{The kinematics of the nebular gas in 30~Doradus}

The large angular size of 30~Doradus (henceforth 30Dor) has made it difficult to fully map the kinematics of the nebular gas. \cite{smiwes} used a single channel Fabry-Perot (FP) interferometer to observe a number of positions in 30Dor with spatial resolutions of $13''; 30"; 120''$.  \cite{melnickPT} (reproduced in \citealt{MTMGP}) followed-up on these observations and showed that the structure function of turbulence in 30Dor was flat, albeit with a spatial resolution of only $13'' = 3.25pc$ . Long before that, however, \cite{feast61} had already found that there was very little structure in the turbulence of the nebula. \cite{chuken} improved upon these previous studies by dissecting 30Dor using long-slit Echelle spectrograms that provided a significantly better spatial coverage, particularly in the outer regions of the nebula. They found that the velocity field of the nebula is dominated by expanding shells of a wide range of diameters, and concluded that the supersonic velocity-width of the integrated H$\alpha$ profile was by and large the result of these expansion motions, although the shells accounted for only about 50\% of the kinetic energy of the gas.  They also dismissed the contribution of gravity in stirring-up the gas kinematics.

\cite{mtt} refined the observations of \cite{chuken} using higher spatial and spectral resolutions, albeit at only one slit position. They confirmed that the supersonic H$\alpha$ profiles are resolved into multiple components, but at all positions along the slit they found a very broad unresolved component of unknown origin. More recently,  \cite{torres} used the multiplexing capabilty of FLAMES/GIRAFFE on the VLT to address some of the issues raised by \cite{chuken} and \cite{mtt}.  While they largely confirmed the findings of these previous studies, they questioned the reality of the broad unresolved component identified by \cite{mtt}.

{ For the ensuing discussion it is important to recall that several phases of the interstellar medium coexist within 30~Doradus: It seems to contain significant amounts of molecular and atomic Hydrogen (\citealt{ox} and references therein); there is the ionized gas at $10^4$K, which we will call the {\em nebular} gas; and there is a very hot X-ray emitting medium at $10^7$K that we will call the hot-gas. Here, when we talk about supersonic motions we are referring to the gas at $10^4$K and a sound speed of $c_s=9.1$~\kms for hydrogen.
  
%In section 2.1, it starts with (1.) by saying that at all positions in 30 Dor, high resolution shows only subsonic line profiles. Then point (2.) says the central regions in 30 Dor also have broad wings: are those resolved into subsonic pieces?  They must be if (1.) is correct. This should be clarified. Following point (3.), it states: "these results lead to the strong conclusion that in 30Dor the supersonic velocity width of the integrated emission-line profiles is the result of the motion of macroscopic parcels of gas propelled by the combined action of gravity and cluster winds." But there is no indication from these statements that gravity is involved.

%The last sentence in section 2.1 says: "within these macroscopic parcels of gas the turbulence is subsonic and uniform." What does "uniform" mean?  Can the motions be random? Does it mean equally turbulent all over the nebula?
 
\subsection{30 Doradus in FLAMES}
\subsubsection{Overall Nebula}

\cite{torres} raised doubts about the reality of the broad component of the integrated \halfa profile of 30Dor identified by \cite{mtt} , so it seems relevant to begin our analysis by revisiting this issue in order to set the ensuing discussion on self-consistent grounds.  We used a subset of the data discussed by \cite{torres} that we downloaded pipeline-processed from the ESO Archive (Phase3). We discarded from this dataset all points showing discernible contamination of the nebular profiles by stellar features, leaving a total of 1668 individual spectra in our full sample. 

Figure~\ref{fig0} shows the integrated profiles determined from these data in two different ways. The top panel shows the average of the 1668 individual FLAMES/GIRAFFE spectra.  \cite{torres} (see also our Figure~\ref{fig100}) provide graphical illustrations of how the data points are spatially distributed in the nebula showing that the central regions are more densely sampled than the outskirts, mostly by the {\it VLT Flames Tarantula Survey} observations \citep{evans}.  Thus, in order to check for effects that this inhomogeneous sampling may introduce,  the lower panel shows the integrated profile using only the data from the uniform grid of \cite{torres} for a total of 879 points.

\begin{figure}[h!]
\vspace{3.6cm}
\includegraphics[height=8cm,trim= 1.cm 0 8cm 10cm]{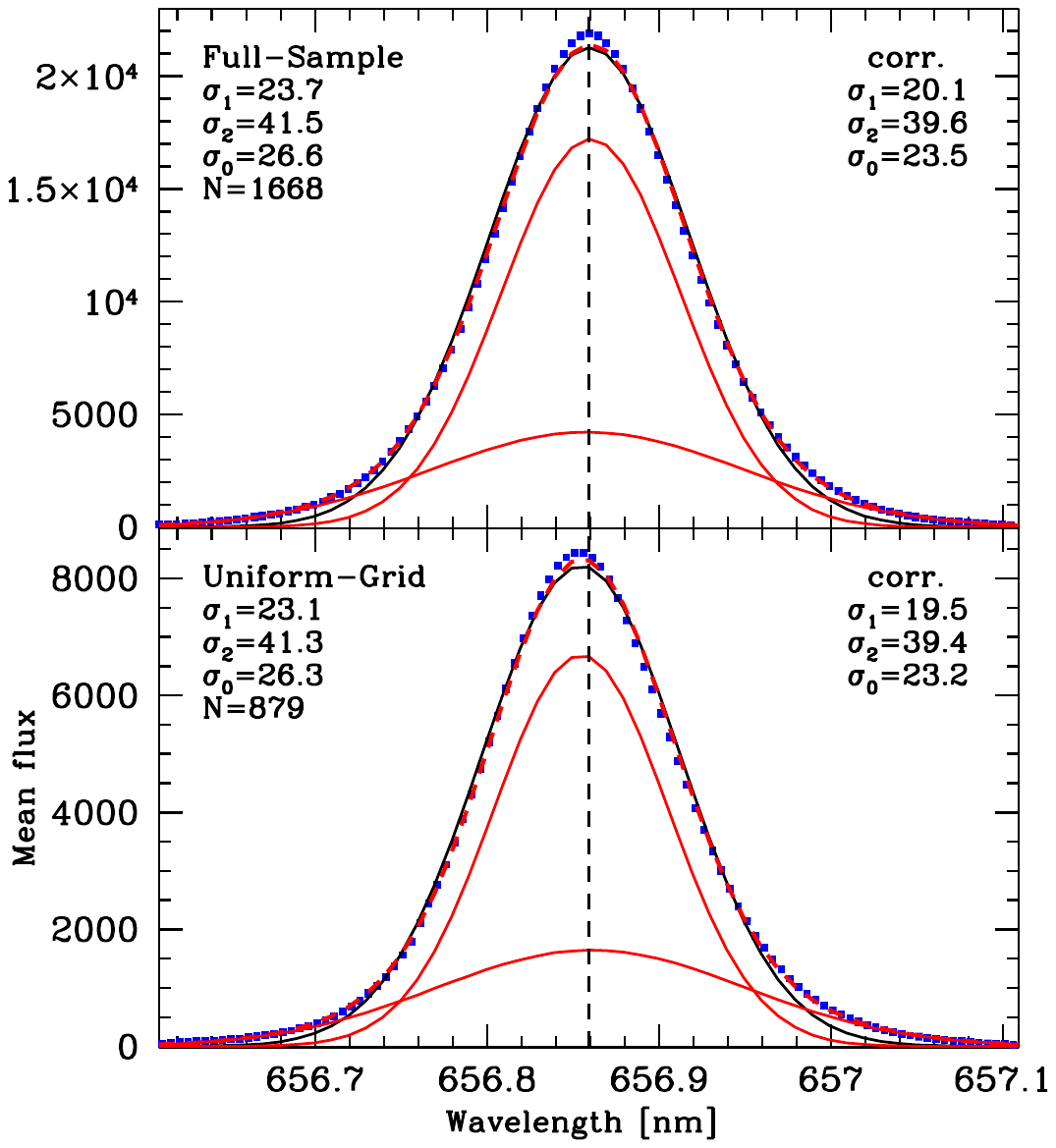} 
\vspace{-2.5cm}
\caption {{\bf Top}. Integrated H$\alpha$ profile of 30~Doradus using all the available data (1668 points). Two Gaussian components are required to fit the core and the wings of the line shown by the thin red lines. The data are shown as blue points and the sum of the two Gaussians by the red dashed lines. The legend reports the measured (left) and corrected for instrumental and thermal broadening (right) widths ($\sigma$) of each component plus the width from a single Gaussian fit ($\sigma_0$ shown by the black lines).  {\bf Bottom}. Same as top, but using only 879 points from the uniform grid of \cite{torres}. The heliocentric radial velocity of 30Dor obtained from the full-sample is $V_{30Dor}=264.8$~\kms for an H$\alpha$ rest wavelength of 656.2804nm.}
\label{fig0}
\end{figure}

Our Gaussian fits for both the Full-Sample and the Uniform-Grid profiles are surprisingly similar given the differences in the two samples. The widths of the profiles also agree with those of \cite{mtt} and \cite{torres} for the core of the line, but the broad component of \cite{torres},  $\sigma_2=49.1$~\kms, is significantly broader than ours ($\sigma_2=41.3$~\kms; lower panel)  {\em basically for the same data}. The discrepancy is much larger than the formal fitting errors ($<1$~\kms), but it is well known that the parameters of multi-Gaussian fits are rather sensitive to the statistical errors of the data, which probably explains the discrepancy. We used the errors provided by the ESO Phase3 pipeline propagated through the various manipulations of the data, but we ignore how \cite{torres} estimated the corresponding errors. \cite{mtt} fitted two Gaussians to the total profile of \cite{chuken} and found $\sigma_1=26$~\kms\ and $\sigma_2=44$~\kms, without including errors.  Notice that two Gaussians do not fit the center of the line for which a third component would be required.  

\subsubsection{Radial bins}

 {The outer regions of the nebula have been characterized by \cite{chuken}, so here we will concentrate in the inner ($r<40$pc) part of 30Dor that contains the bulk of the ionizing stars and has a complex structure that was not well covered in the study of \cite{chuken}. 

To begin, we present in Figure~\ref{fig1} profiles integrated over 7 radial rings of radii indicated in the legend. For each ring, the figure shows multiple Gaussian fits with the minimum number of components required to reproduce the observations. 

 \begin{figure}[h!]
\vspace{5cm}
\includegraphics[height=12cm,trim= 3.8cm -1 7cm 10cm]{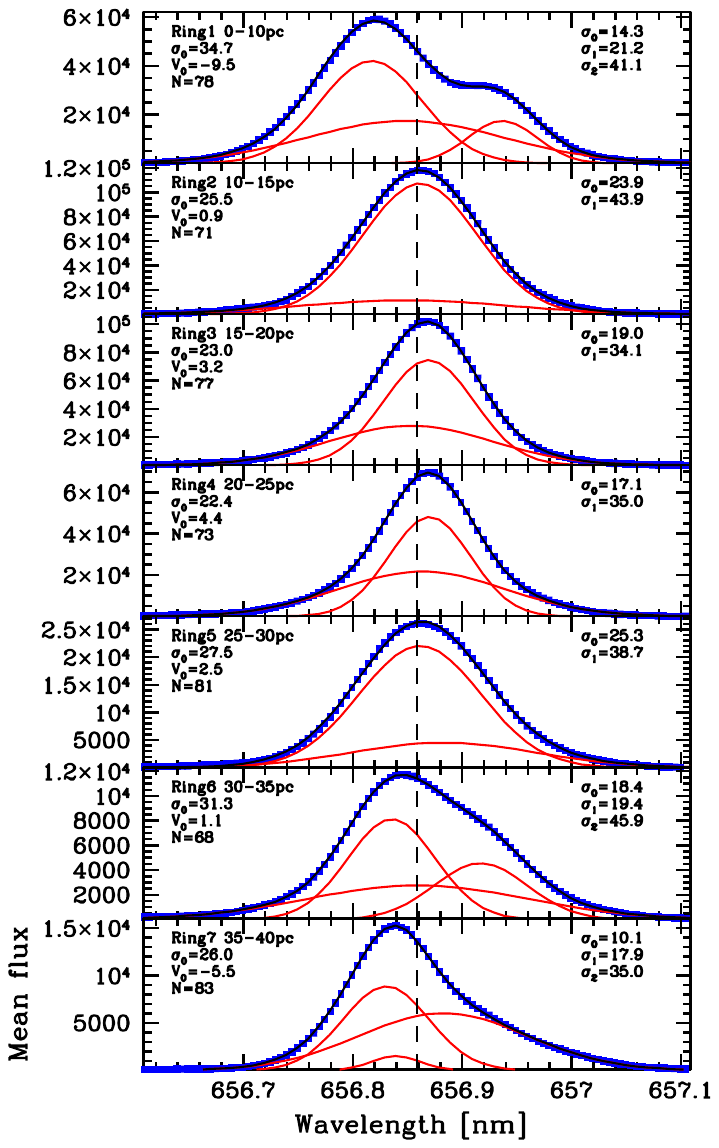} 
\vspace{-3.7cm}
\caption {Integrated profiles of the core of 30Dor ($r<40pc$) and 7 radial bins { identified in the figure legend that quotes the inner and outer radii of each ring in pc. The legend also gives in the right side widths of the multi-Gaussian fits of each ring and on the left, denoted as 0, the width and radial velocity of the single Gaussian fit (that is not plotted) relative to the integrated velocity of the nebular core that defines our rest-frame for 30Dor. }}
\label{fig1}
\end{figure}

We will insist throughout this work that multiple-Gaussian fits are seldom unique being increasingly sensitive to data errors and to the initial guesses as the  number of Gaussian components increase, so we must be conservative when interpreting the results from these fits.   Nevertheless, Fig.~\ref{fig1} shows several robust features relevant to the present study. One is that the innermost ring ($r<10$pc) shows two intense peaks that tell us that the kinematics of the very center of the nebula is dominated by a massive wind that is blowing away the gas at high speeds.  Despite this raging wind, however,  there is still plenty of gas at the center of 30Dor as indicated by the intensity scale.   

A second feature is that there is a dramatic change in the kinematics at $r=25$pc. Beyond this radius, the profiles become broader and increasingly show the emergence of a second redshifted component. Thus, beyond 25pc we see the onset of the "large expanding-shells" regime. A third feature is that the surface brightness drops dramatically as the structure of the gas becomes increasingly dominated by the large shells described by \cite{chuken}.} \\

%3, Point 1 above holds also through the first 30 pc, although both kind of lines present a lower intensity as one moves outwards.
%Section 3.1.2 mentions 1869 profiles for 30 Dor but section 3.1.1 mentions 1668. What are they different?

%Section 3.2 says: "The strongest profiles, located predominantly within the core of the nebula, tend to be narrower, while the broadest profiles tend to be located in the fainter outer regions" whereas Section 3.1.2 says: "The figure shows a clear tendency of the profiles becoming narrower outwards from the core." These statements, both for 30 Dor, appear to be contradictory if we look at position versus profile width. This should be explained better.

%Section 3.2 says: "Therefore we used a simple algorithm described below to separate single and multiple profiles." But I thought all profiles were multiple, with many separate subsonic components. Are the profiles discussed here too poor resolution to see subsonic substructure? The instrumental resolution is said to be 7.8 km/s (section 3.2). Is that small enough to see the separate components? It also says: "while the profiles of the brighter more central regions tend to be single and narrower" - does that mean subsonic? The authors should explain better the context of this section and what they are doing and how it relates to the rest of the paper.
 
\subsection{Profile taxonomy}
\label{singdou}

The atlas of gray-scale spectrograms of \cite{chuken} shows that virtually everywhere in the nebula the H$\alpha$ profiles tend to break-up into multiple components. However, the high contrast of the atlas prints makes it difficult to precisely discern the profile shapes. Thus, we were surprised to discover, upon visual examination of our 1668  H$\alpha$ profiles in the computer screen, that the majority of the profiles can  actually  be divided into three well defined groups illustrated schematically by Figure~\ref{fig3}.

\begin{figure}[h!]
\vspace{0.3cm}
\hspace{-0.55cm}\includegraphics[height=4.8cm]{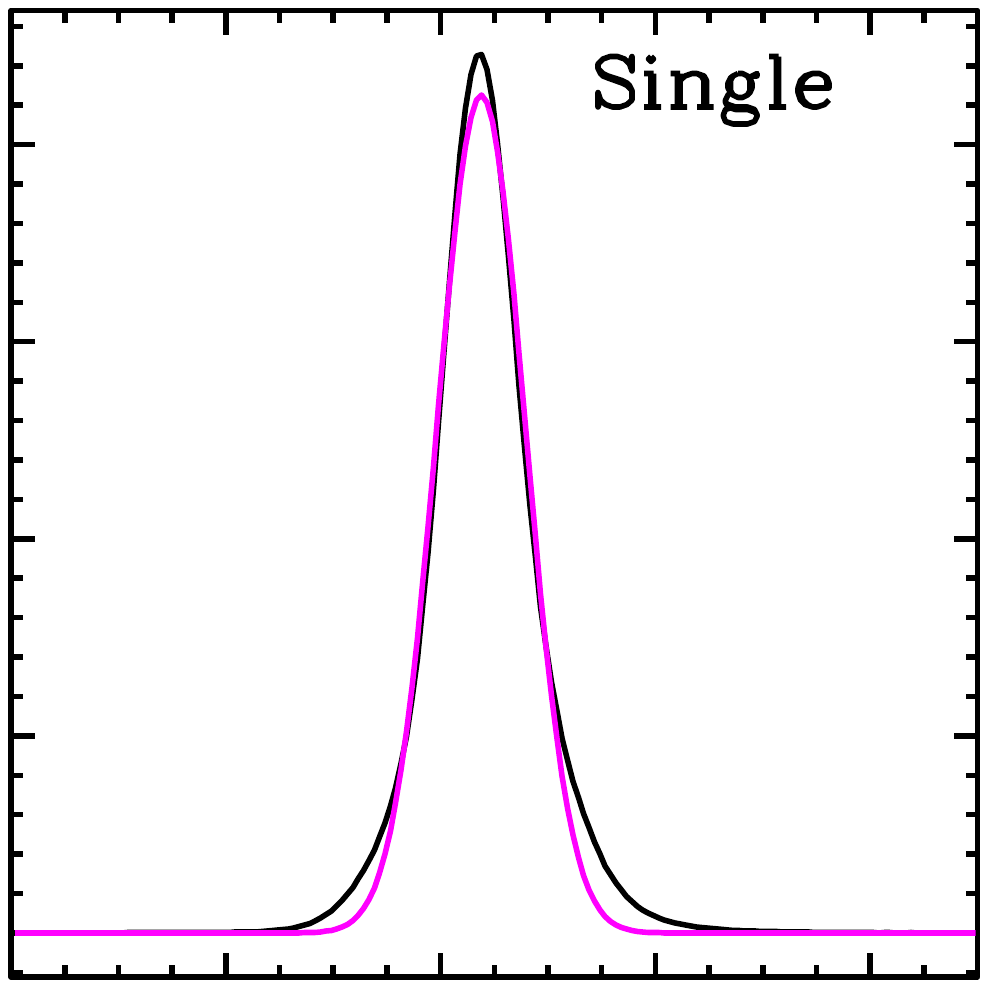}\hspace{-0.79cm}\includegraphics[height=4.8cm]{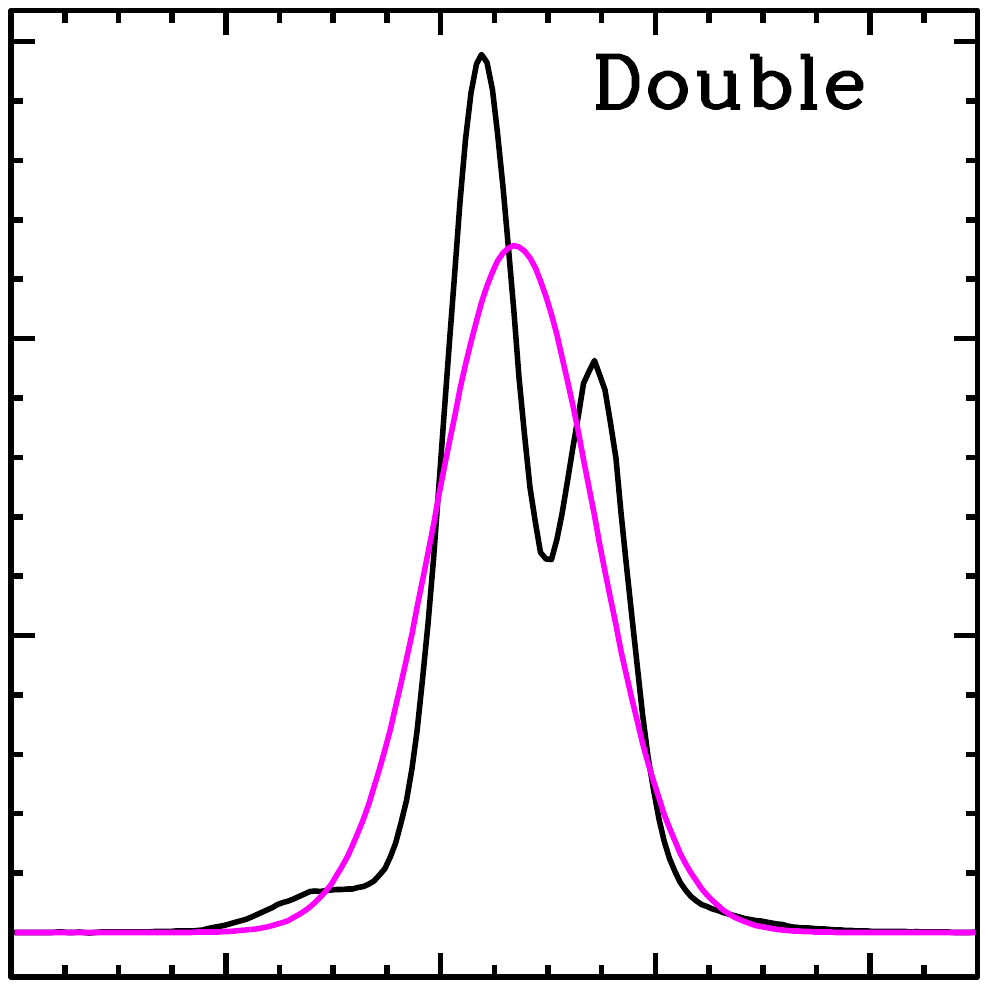}\hspace{-0.79cm}\includegraphics[height=4.8cm]{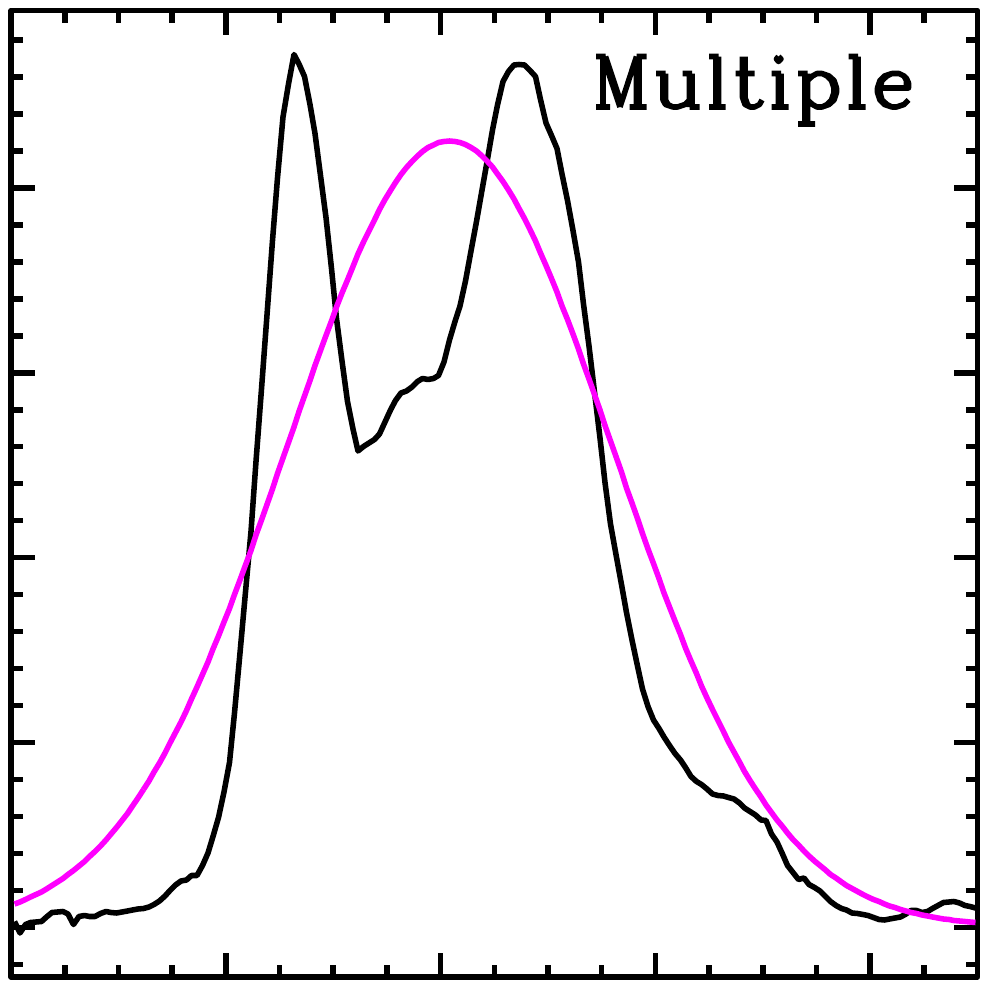}
\vspace{-1.4cm}
\caption {Cartoon illustrating the three typical shapes of our 1668 profiles. The intensity ratios of the double and multiple components vary widely as discussed in the text. The purple lines show single Gaussian fits to the particular profiles illustrated in the cartoons. }
\label{fig3}
\end{figure}

The vast majority of the profiles are either single or double, although multiple profiles also contribute a fair fraction. We emphasize that we here are describing the \halfa profiles. We will discuss the forbidden lines in the next section. Like the integrated profile, single profiles have Gaussian cores and broad non-Gaussian wings. Double and multiple profiles are characterized for having  two or more smooth components of comparable (within factors of a few) peak intensities that are well separated in radial velocity. The radial velocity of the stronger component generally appears to be closer to zero, whereas the weaker peak is either redshifted or blue shifted relative to the stronger component.

We mentioned at the beginning that \cite{chuken} concluded that 50\% of the kinetic energy of the gas is in the expanding shells and 50\% in some form of unspecified
turbulence. Assuming that double and multiple represent pure expansion, the single profiles must convey information about that "unspecified" form of turbulence. Therefore, it is important to quantify the distinction between single and multiple profiles. }
 
We used an algorithm based on first and second derivatives to detect multiple peaks using the $S/N$ of the continuum to weedout spurious peaks. Most of the profiles show some degree of low-level asymmetry, which does not make them multiple in the sense of having more than two peaks of comparable intensities. Therefore we introduced a finer distinction in order to separate single from multiple profiles. We defined as "double" profiles where the ratio of peak-intensities between the first (strongest) and second components $P_2/P_1>0.1$, and "multiple" profiles having three or more components with $P_3/P_1>0.15$. 

The peak finding algorithm  yields 1067 singles; 453 doubles and 148 multiples, confirming our initial impression that multiple profiles are less abundant but still numerous. Visual examination revealed that a good fraction of singles were actually doubles where the radial velocities of the peaks are too close to be detected by our peak finding algorithm.

Since the H$\alpha$ profiles are blurred by thermal motions that hide components with a small difference in radial velocity, we refined our separation of single and multiple profiles using the lines of [NII] and [SII] that have much smaller thermal broadenings (2.4 and 1.6~\kms respectively). This "solution", however, turned out to be complicated as discussed below.

\subsection{[NII] and [SII] kinematics}
\label{rafa}

\begin{figure}
\vspace{1cm}
\includegraphics[height=7cm, trim= 1cm 0 0.8cm 12cm]{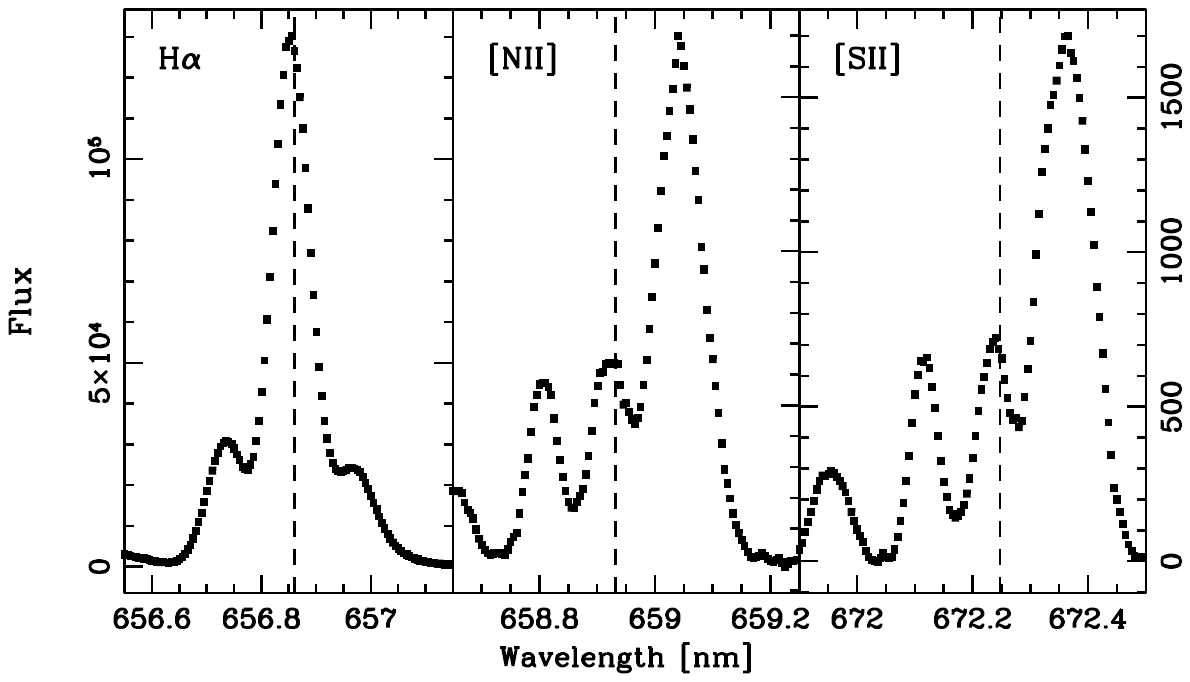}
 \vspace{-2.5cm}
\caption {\small The profiles shapes of \halfa are generally different from those of the forbidden lines. The vertical dashed lines correspond to the wavelengths at the heliocentric radial velocity of 30Dor from the sum of the single \halfa\ profiles $V_{30Dor}=264.4$~\kms.   }
\label{nalfa}
\end{figure}

Using the [NII] and [SII] lines (henceforth the forbidden lines) to characterize the nebular gas kinematics has two problems. The first most obvious one is that the lines are much weaker than \halfa and often their S/N is too low for a reliable characterization of the profiles. The second, less obvious problem, is that in general the peak intensities of the multiple components of \halfa differ from those of the forbidden lines, as in the example shown in Figure~\ref{nalfa}. 

It is not appropriate, therefore, to use the morphology of the forbidden lines to characterize the kinematics of \halfa (and viceversa).  So we only used the forbidden lines to flag single and multiple profiles, and we searched for multiple components in all profiles thus flagged by subtracting single Gaussian fits from the observed \halfa profiles and fitting multiple Gaussians, using the positions of the residual peaks as initial guesses, to measure the peak intensities. 

In this way we found that 599 of the 1067 lines classified as "single" in H$\alpha$ by the peak finding algorithm were double in [NII] and/or [SII] and confirmed to have $P_2/P_1>0.1$ at \halfa, while 30 profiles previously classified as double were confirmed to be multiple ($P_3/P_1>0.15$).

Relevant statistics from the three families of \halfa profiles are presented in Table~\ref{tab3} that lists average radial velocities of the strongest components $<V_1>$; the average peak intensity ratios; and the average velocity-widths from single-Gaussian fits (uncorrected for instrumental and thermal broadening; $\langle\sigma\rangle$). The $\pm$ values give the rms dispersions from the means.  \\

\begin{table}[!ht]
\tabcolsep 1.5mm
\vspace{0cm} 
\tiny
\caption{\bf Taxonomy of H$\alpha$ profiles in 30~Doradus}
\begin{tabular}{ l c c  c c c  }
\hline\hline
\hspace{0.5cm}                                                                                                                           & N       & $<V_1>$              & $ <P_2/P_1>$   &  $<P_3/P_1>$  &  $ <\sigma>$  \\ \hline 
\vspace{-0.5cm}
\hspace{-0.15cm}\parbox[c]{1em}{\includegraphics[width=1cm]{rocs3.pdf}}   \hspace{0.4cm}  & 468    & $265.4\pm11.6$   &  ----                    &   ----                  & $17.3\pm3.4$  \\
\vspace{-0.5cm}
\hspace{-0.15cm}\parbox[c]{1em}{\includegraphics[width=1cm]{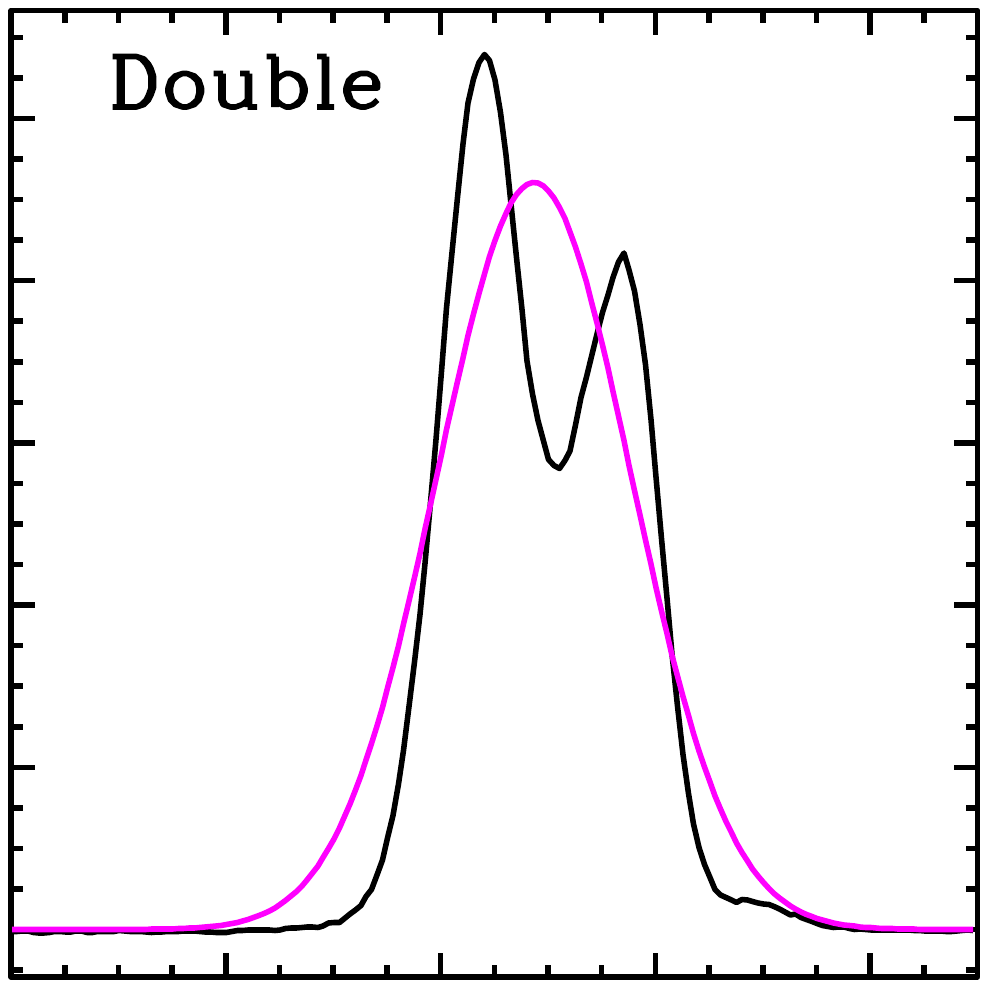}}   \hspace{0.4cm} & 1043 & $263.8\pm18.6$   &  $0.36\pm0.27$ & ---                     & $24.6\pm8.0$  \\ 
\vspace{-0.3cm}
\hspace{-0.15cm}\parbox[c]{1em}{\includegraphics[width=1cm]{rocs41.pdf}} \hspace{0.4cm} & 157   & $265.6\pm28.6$   &  $0.58\pm0.30$ &  $0.41\pm0.27$  & $36.1\pm18.4$  \\ \hline
\end{tabular}
\label{tab2}
\end{table}
\vspace{-0.05cm}
 
\begin{figure}[h]
\vspace{6cm}
\includegraphics[height=4.5cm, trim= 0 -1.5cm 0.8cm 18cm]{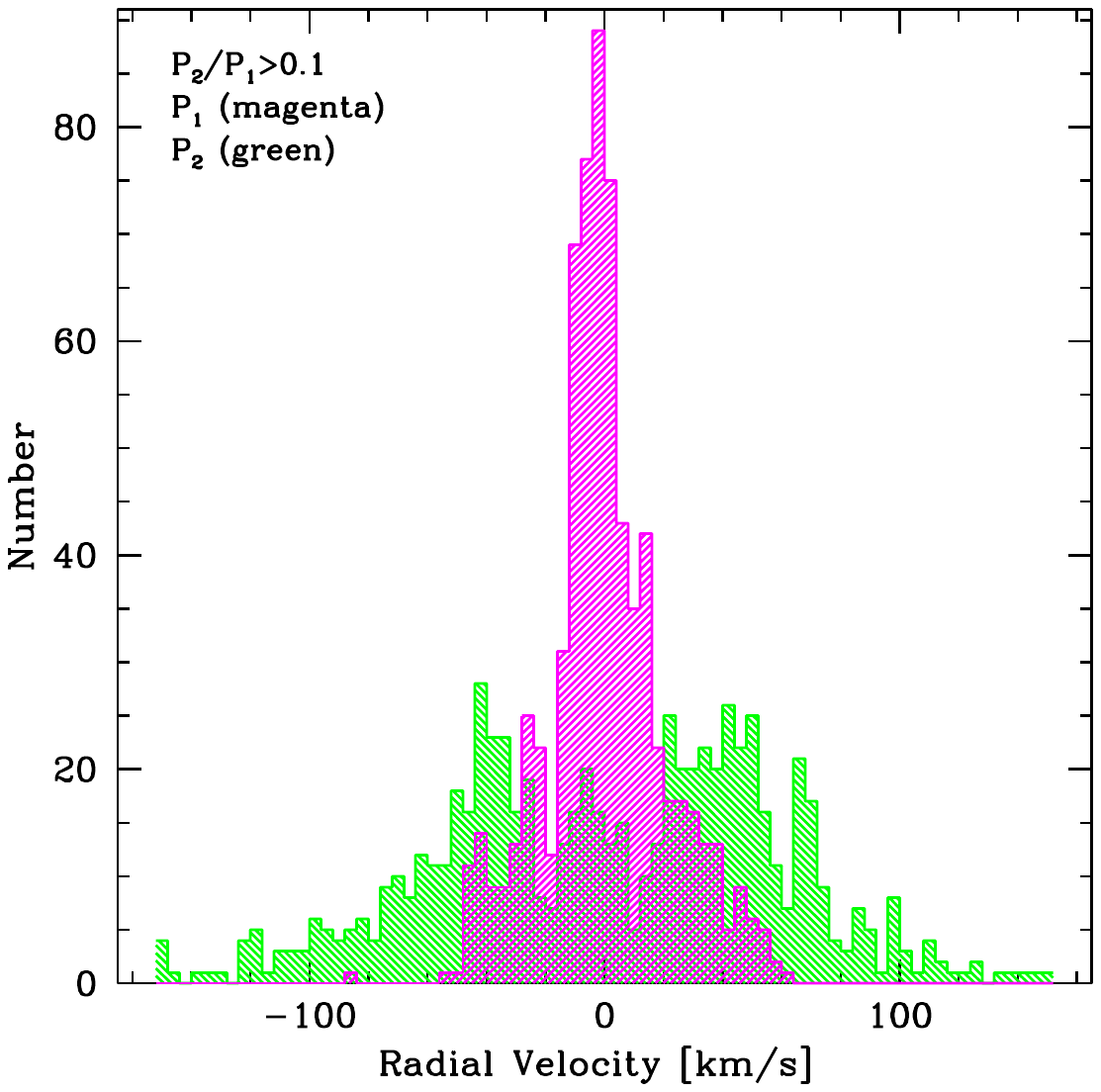}
\vspace{-2.8cm}
\caption{ Distribution of radial velocities relative to the rest frame of 30Dor of the two components of the double profiles. $P_1$ (magenta) represents the component with highest peak intensity and $P_2$ (green) the weaker component.}
\label{fig5}
\end{figure}

Using the partition of single and double profiles we can now quantify our second visual impression that the stronger component is generally at rest in the frame of the nebula { while the weaker component is either redshifted or blue shifted relative to the strong component.} Figure~\ref{fig5} shows the histograms of radial velocities of the two components of our "double" profiles. As in Table~\ref{tab2}, $P_1$ corresponds to the strongest peak and $P_2$ to the weaker one.

It should be obvious from these distributions and the data of Table~\ref{tab2} that the integrated profile of 30Dor can be deconstructed into a sum of Gaussians representing  the components of the double peaks plus the single peaks and their rms radial velocity dispersion. 

\begin{figure*}
\hspace{0.5cm}\includegraphics[height=10cm]{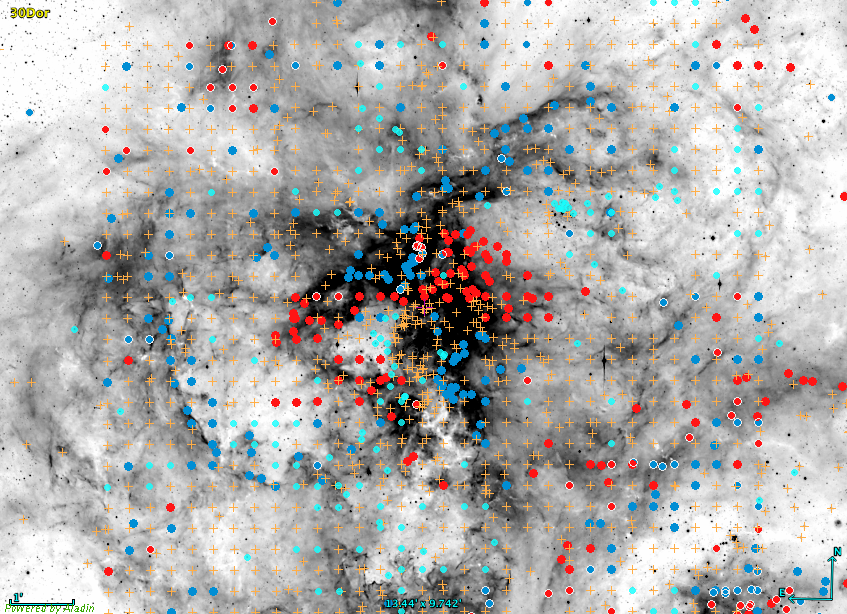}
\vspace{-0cm}
\caption {\small Public H$\alpha$ image of 30Dor taken with the Wide Field Imager on the 2.2m telescope of La Silla. The dots and crosses show the positions of the three families of profiles discussed in the text color-coded as follows: blue for blue-shifted single profiles; red red-shifted single profiles; yellow crosses doubles; and cyan dots multiple profiles. The red and blue dots with white rims have radial velocities close to zero ($|V|<2.5$~\kms). The lower-right corner (SW) of the image portrays the slow expanding shell of the supernova remnant 30DorB. The scale shown at the lower-left corresponds to 15pc.}
\label{fig100}
\end{figure*}

Finally, let's look at the spatial distribution of our three families of profiles. This is presented in Figure~\ref{fig100} where we have color coded the positions according to morphology.  Doubles tend to concentrate inside the large expanding shells; singles tend to delineate the high surface brightness borders of these shells; while multiples share the distribution of doubles.

The image illustrates that the profiles inside the large Chu-Kennicutt bubbles are double or multiple, while at the edges of these giant shells the profiles are mostly single. We have color coded the single profiles according to their peak radial velocities relative to the systemic velocity of 30Dor (265.4\kms) from our \halfa data. The edge of Chu-Kennicutt\#2, that has been very well mapped in search for photodissociation regions, is seen to be blue-shifted indicating that the shell is expanding towards us, while the inside shell that was identified by \cite{chuken} but not numbered is completely delineated by red dots.

The single profiles, therefore, delineate the edges of the expanding shells as well as the dense filaments and loops that characterize the core of the nebula, and which will be described in detail below. Notice that the blue and the red dots are rarely mixed showing that the densest parts of the nebula as a whole are expanding.

\subsection{Single Profile Kinematics} 
 
Assuming that the kinematics of the nebular gas represented by double and multiple profiles is dominated by expanding motions, single profiles should provide us with information about the 50\% of the total kinetic energy of unknown forms of turbulence mentioned by \cite{chuken}.  We have just seen that the radial velocities of the single profiles are dominated by expansion motions, but the individual profiles also have supersonic widths.

Table~\ref{tab3} summarizes the average intrinsic width (i.e. corrected for instrumental and thermal broadening) and heliocentric radial velocity of the single profiles for \halfa and for the forbidden lines.

\begin{table}[!h]
\vspace{0cm} 
\caption{\bf Kinematics from single profiles}
\tabcolsep 1.5mm
\small
\begin{tabular}{ c c c c}
\hline\hline
Line                   & N               &  $<\sigma>$        & $<V_{rad}>$          \\ 
                          &                  &  km/s                   & km/s                       \\ \hline 
 H$\alpha$        & 478           & $11.9\pm3.4$      & $265.4\pm11.6$     \\ 
 $\rm [NII]$        & 474           & $10.8\pm3.6$      & $262.6\pm13.0$     \\  
 $\rm [SII]$        & $256^a$   & $10.9\pm3.2$      & $263.3\pm9.4$       \\ \hline
\end{tabular}
\tablefoot{\tablefoottext{a}{\small The wavelength range of the uniform grid of \cite{torres} does not reach [SII]}}
\label{tab3}
\end{table}

The table shows that the forbidden lines are on average $\sim1.3$~\kms\  {\em narrower} than H$\alpha$.  { Figure~\ref{fig6} shows the stacking of the single [NII] and H$\alpha$ profiles after rebinning the data to $z=0$. We have repeatedly whined about the dangers of multi-Gaussian fits, but in the case of Fig.~\ref{fig6} a {\em minimum} of three components is required to fit the core and the wings of the lines. In fact there are good physical reasons to fit three components; we have already encountered two: $\sigma_1$, the main peak, and $\sigma_3$, the broad unresolved component that pervades the nebula. The third component ($\sigma_2$) corresponds to the sum of all the low-intensity components that do not meet our $P_2/P_1>0.1$ and/or $P_3/P_1>0.15$ conditions.   
 
\begin{figure}[h]
\vspace{6cm}
\includegraphics[height=3.5cm, trim= 1cm 0 9cm 20cm]{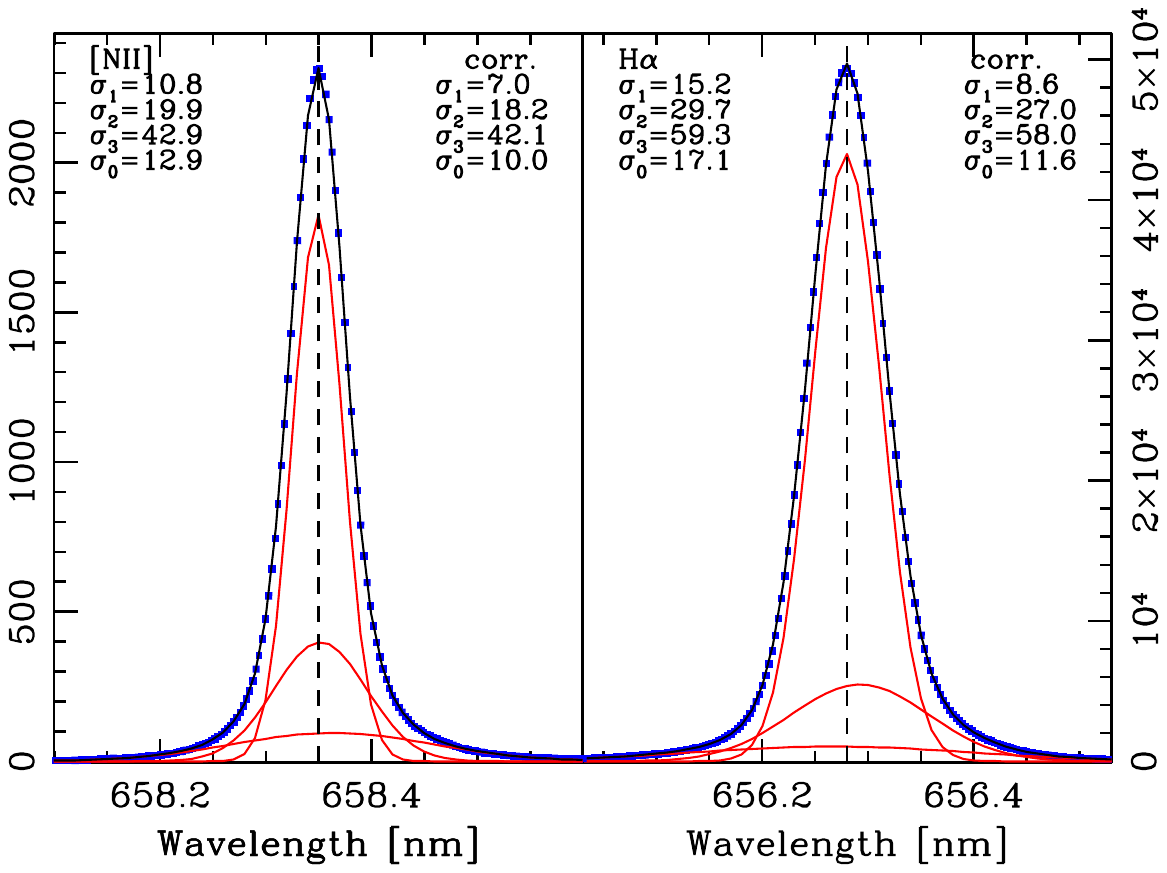}
\vspace{-2.6cm}
\caption {{\bf Left.} Stacked [NII] single profiles rebined to zero redshift using a wavelength of 656.2804 for H$\alpha$. The red lines show the three Gaussian components necessary to fit the data. The blue dots show the data, and the black line the overall fit. The legends give the observed width of each component top left, and the widths corrected for instrumental and thermal broadening on the top right. $\sigma_0$ corresponds to a single-Gaussian fit. {\bf Right.} Stack of H$\alpha$ profiles.}
\label{fig6}
\end{figure}
 
A second interesting feature of Fig.~\ref{fig6} is that, as already shown in Table~\ref{tab3}, the H$\alpha$ profile is broader than [NII] by 1.6~\kms. The difference is somewhat larger than the value of 1.3~\kms\ from the table that assigns the same weight to all profiles independently of peak intensity, so the agreement is actually quite good. 

We verified that the systematic difference in the line-widths is not due to a geometrical effect (see Appendix~\ref{ap1}). In fact, \cite{hip86} found a similar systematic difference between [OIII]500.7nm  and H$\alpha$\ for the integrated profiles of a small sample of Giant HII Regions and HII Galaxies.  This difference has been  confirmed by several authors notably by \cite{mel2017} for a substantial sample of HII Galaxies.  A similar effect was also observed in the Orion Nebula by \cite{tedoi} but there is no consensus in the literature as to the origin of the discrepancy. { Notice, however, that while the {\em intrinsic} width of H$\alpha$ ($\sigma_1$) is the same as the sound speed for hydrogen at $10^4$K, the [NII] profiles are highly supersonic, suggesting that perhaps they could be resolved into even more components.}

We also measured the intrinsic widths of each of the individual components of double H$\alpha$ profiles {\bf - taken separately -} using two Gaussians to fit the doublet plus a third Gaussian to fit the extended wings, using an automated procedure. We find an average width of $11.7\pm4.7$~\kms for these components, comparable to the values given above for the average width of single profiles ($11.9\pm3.4$\kms).

%The authors say: "The difference in the width of [NII] and Halpha, therefore, is ultimately due to the thermal broadening of the Balmer lines and not to the astrophysics of ionized nebulae." but I thought thermal broadening was removed from both profiles (in addition to instrumental effects). The figure 5 caption says: "The velocity width of each profiles corrected for instrumental and thermal motions are shown in the legend." DONE

%By the time the authors say the following, I am really confused: "The main result from our analysis of the [NII] and [SII] lines is that (after correction for thermal and instrumental broadening) the bona-fide single profiles in 30 Doradus still have supersonic velocity widths. The same conclusion applies for the individual components of the double profiles." I thought the discussion in the introduction was about subsonic components of supersonic unresolved profiles. The logic of the paper is hard to follow since the conclusion at this point is apparently just the opposite of what the introduction says (and also what the conclusion says). Also the authors discuss the main line component and the wings and, as mentioned above, it is not clear in the introduction whether the wings were considered before to be also composed of subsonic components. At this point in the paper it seems like the authors are saying that both the core and the wings are not composed of subsonic compone!nts. Is that right? DONE

\subsection{The [OIII] profiles with HARPS}

{We speculated above that a possible explanation for the reason the intrinsic width of single [NII] profiles is highly supersonic while H$\alpha$ is not, could be that at higher resolution the  profiles would be resolved into multiple narrower components. We therefore observed two characteristic regions of 30Dor with HARPS: one at the very center  (R136),  and the second at the position of Filament~F from the NTT observations of \cite{mtt}, located 8pc due west of R136. The original purpose of this pilot-project was precisely to resolve the profiles further, and to verify if the forbidden [OIII] lines also showed the broad unresolved component seen by \cite{mtt} at H$\alpha$.}

HARPS provides a spectral resolution of R=100000 covering the spectral range between 370nm and 690nm with a spatial resolution, determined by the entrance fiber, of $3''$. 
Figure~\ref{fig7} shows the HARPS spectrum of the center of 30Dor (R136) that nicely illustrates the extremely broad "feet" of the Balmer lines from the massive WN7h stars in R136 that provide a significant fraction of the total ionizing flux \citep{doran}. Similar features are observed at several positions within the nebula where such massive stars are found \citep{evans}. 

\begin{figure}[h]
\vspace{3.75cm}\includegraphics[height=4.4cm, trim= 1cm 0 0 20cm]{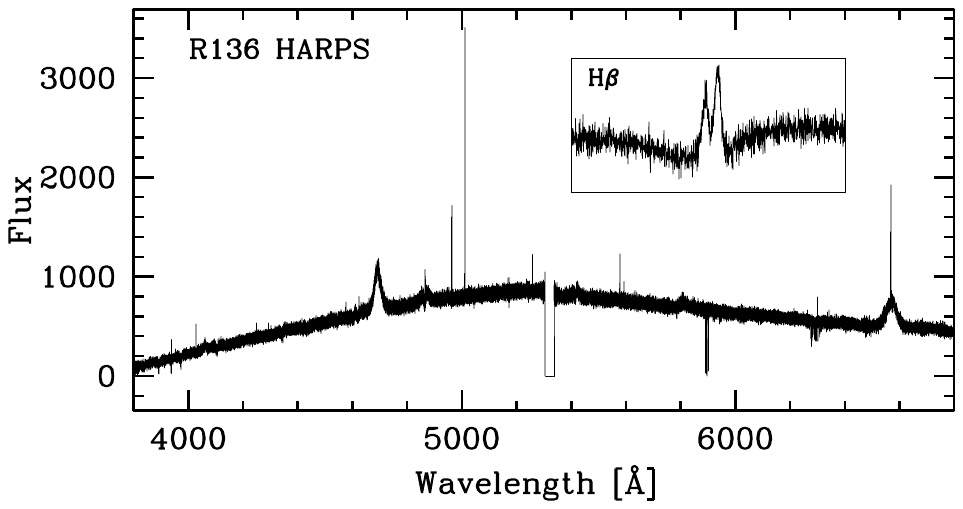}
\vspace{-3.4cm}
\caption {HARPS spectrum of the center of 30 Doradus (R136) showing the broad features of the massive WN7h stars in the core
of its ionizing cluster. The  insert shows the nebular H$\beta$ line superposed on the broad absorption trough of the stellar component.}
\label{fig7}
\end{figure}

Figure~\ref{fig8} presents a comparison between the H$\alpha$ and the [OIII]5007$\AA$~ nebular profiles at the two positions observed with HARPS.
\footnote{In fact HARPS provides two fibers - object/sky - at each pointing, but in our case the sky fibers have significantly lower S/N so we chose not to use them here.}  
 
\begin{figure}[!h] 
\vspace{3.9cm}\includegraphics[height=4.cm, trim= 0.5cm 0 8cm 20cm]{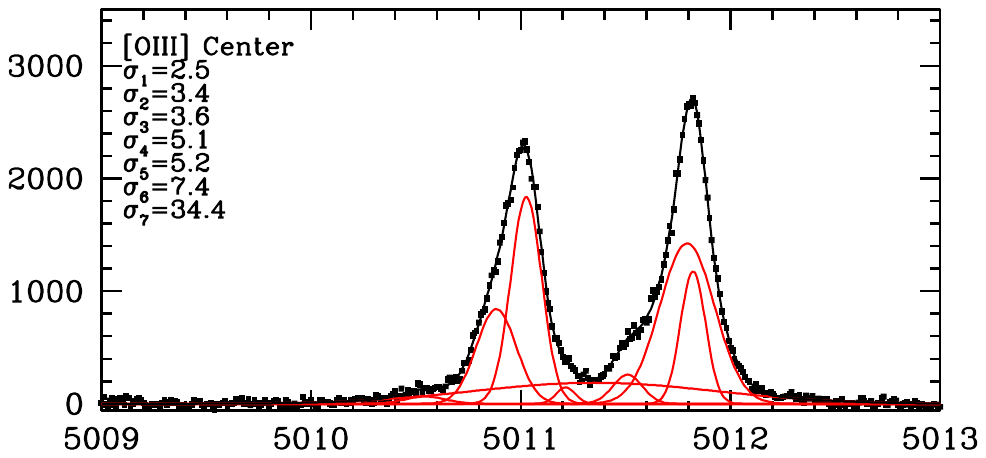} \\
\vspace{0cm}\includegraphics[height=4.cm, trim=0.5cm 0 8cm 20cm]{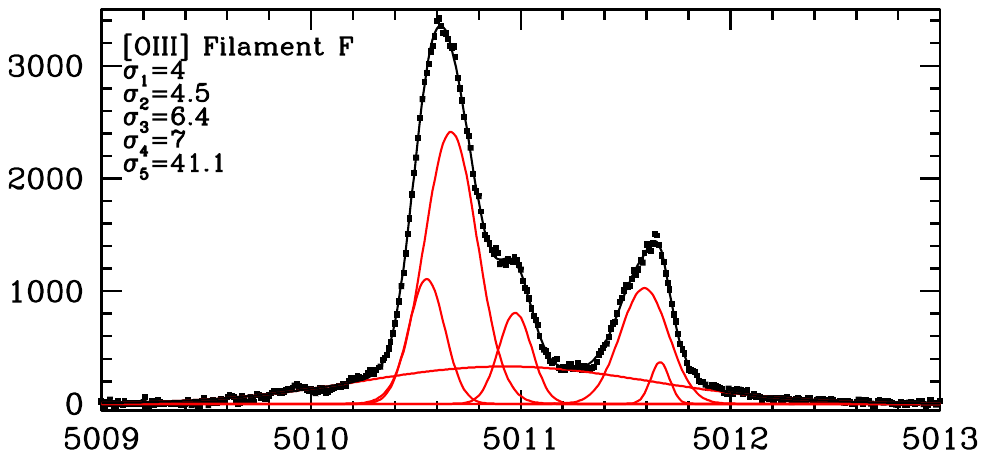} \\
\vspace{-0cm}\includegraphics[height=4.cm, trim=0.5cm 0 8cm 20cm]{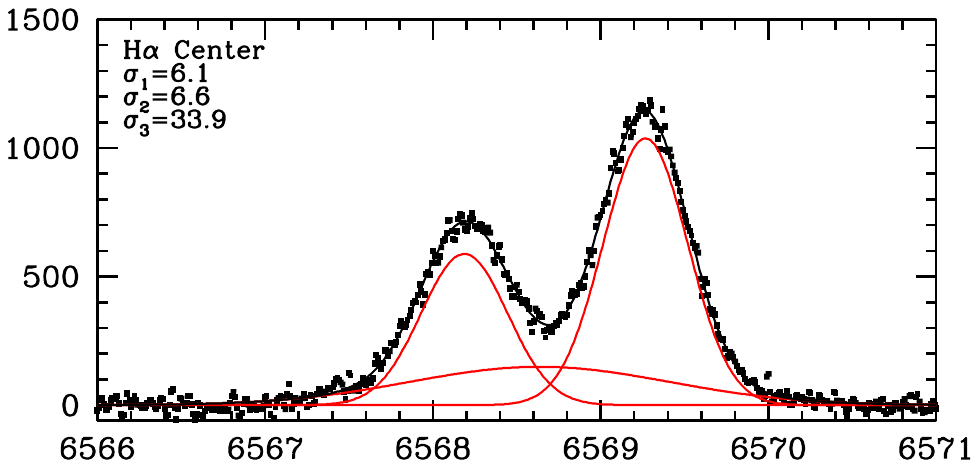} \\
\vspace{0.cm}\includegraphics[height=4.cm, trim=0.5cm 0 8cm 20cm]{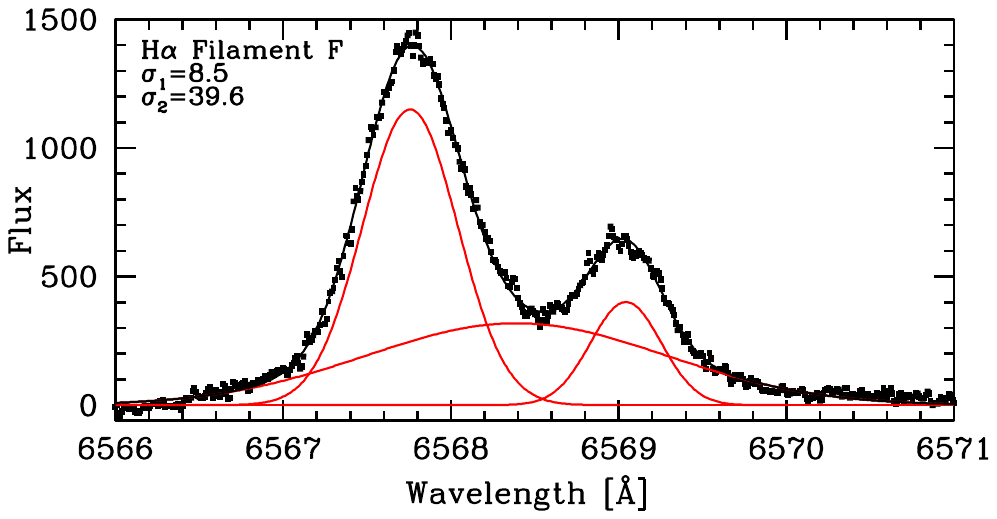}  
\vspace{-2.6cm}
\caption {\small Nebular line-profiles at the two positions observed with HARPS. H$\alpha$ profiles in the bottom panels and [OIII]5007 
at the top. The multi-gaussian fit components are plotted in red and the solid black line shows the overall fit. }
\label{fig8}
\end{figure}

The H$\alpha$ profiles shown in the bottom panels of the figure are double in the sense discussed above: they can be fitted reasonably well using two Gaussian components for the two main peaks plus a third for the broad extended wings. The [OIII] lines, on the other hand, have higher S/N and much smaller thermal broadening (2.3 vs. 9.1~\kms at $10^4$K), allowing to discern a significantly more complex structure at the same positions, which requires at least 7 components, plus one for the broad wings, at the center, and at least 5 components plus the broad wings for Filament~F.  The corrected velocity-widths of the individual components are given in the figure legends.

{ Although the [OIII] line resolves into several components, these are still supersonic at $10^4$K, but  again we must return to the "mantra" of this paper: {\em beware of multi-Gaussian fits}. The solutions shown in Figure~\ref{fig8} result from leaving all parameters free. It is actually possible obtain solutions by forcing the widths of all components to be narrow, say, $\sigma=3$~\kms, albeit only after carefully fine-tuning the initial guesses of all other parameters. So we can neither prove nor rule out that the [OIII] profiles are the sum of many subsonic components, for which we would need much higher S/N data. } 

Nevertheless, our results clearly show that the individual components of double profiles are resolved into a number of narrower components indicating that there is structure in the expanding shells at the spatial resolution of HARPS (0.75pc). The radial-velocity dispersion of these components is "typically" (we only have 4 examples) $\sim 11$~\kms, quite similar to the value measured from the unresolved shells and filaments.

\section{The kinematical core of 30~Doradus}
\subsection{Structure}
{ Figure~\ref{fig1} shows that there are two inflection points in the integrated profiles of the central part of 30Dor: the region $r<10$pc, where the profile is double and the kinematics dominated by the massive cluster wind emerging from the stars in the cluster core (R136); and the region $r>25$pc where expanding motions increasingly dominate the profile shapes. We used finer bins to verify that the latter change is already visible between 25pc and 27pc.}

{ The radial H$\alpha$ brightness distribution, shown in Figure~\ref{fig108}, also shows features (i.e. changes of slope) at $r\sim10$pc and $r=25$pc amid a broad peak that encompasses 33\% of the total H$\alpha$ flux of the nebula and the bulk of the ionizing stars, and where the cluster wind that drives the hierarchy of Chu-Kennicutt shells originates. }

\begin{figure}[h]
\vspace{7cm}
\includegraphics[height=3.3cm, trim= 1.0cm 0 8cm 20cm]{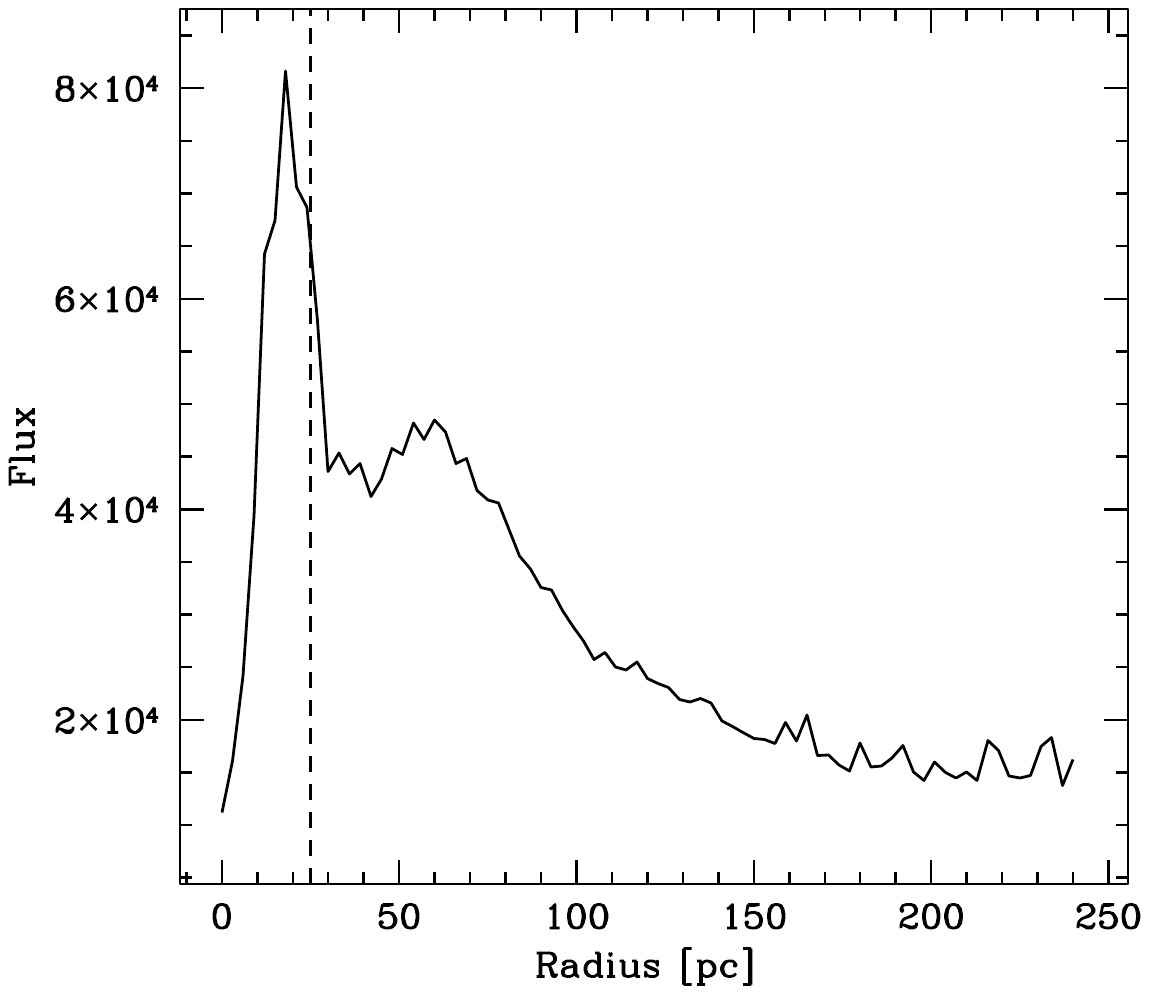}
\vspace{-3cm}
\caption {Radial H$\alpha$ brightness profile of 30Dor obtained integrating the flux in incremental rings of 1pc. The dashed line shows the radius, $r=25$pc, where the kinematics shown in Figure~\ref{fig1} shows a distinct inflection point.}
\label{fig108}
\end{figure}

{ Figure~\ref{fig108a} shows that the width of the integrated profile of the region $r\leq25$pc (henceforth the {\em kinematical core}) is virtually identical to that of the nebula as a whole (the only significant difference being  the width of the broad component).  Therefore, the ratio of kinetic energies of the nebula as a whole and the kinematical core is just the ratio of the \halfa luminosities; the core contains 23\% of the total kinetic energy of the nebular gas in 30Dor.  Perhaps tellingly, the shape of the profile of the kinematical halo ($r>25$pc;  the realm of the large expanding bubbles) is virtually identical to those of the core and of the overall nebula.

\begin{figure}[h]
\vspace{8cm}
\includegraphics[height=3.6cm, trim= 1.0cm 0 8cm 20cm]{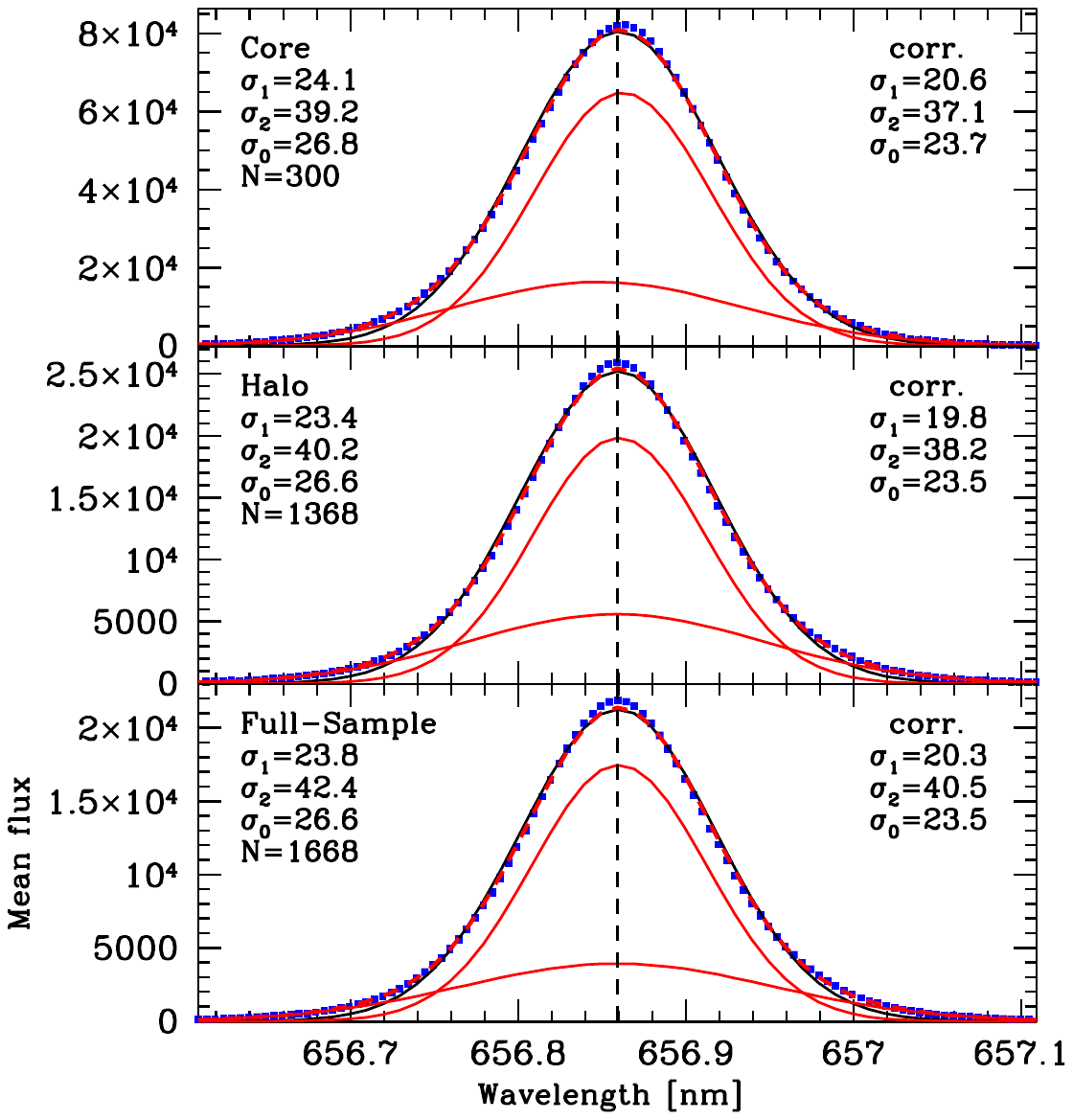}
\vspace{-2cm}
\caption {Comparison between the integrated H$\alpha$ profiles of the whole nebula (bottom) with the profile of the kinematical core ($r\leq25$pc), and the kinematical halo ($r>25$pc). The blue points are the data and thick-dashed red line is the fit of two Gaussian components shown in thin red. The single-Gaussian fit is shown by the black line. The parameters of these components, observed and corrected for instrumental and thermal broadening are shown in the legend. }
\label{fig108a}
\end{figure}

{ Figure~\ref{fig110} presents a zoom to the central 30pc of 30Dor seen in X-rays (right) and H$\alpha$ (left). Overlayed on these images are colored dots coded according to profile shapes and kinematics.}
 
\begin{figure*}
\hspace{0.cm}\includegraphics[height=8.65cm]{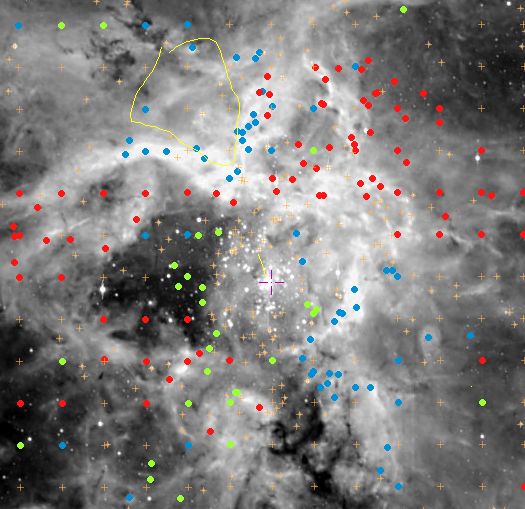}\vspace{0cm}\includegraphics[height=8.65cm]{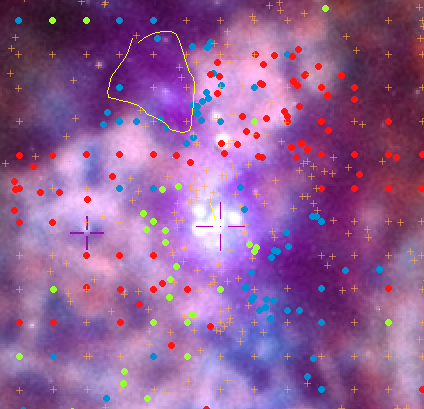}%\vspace{0cm}\includegraphics[height=5cm]{spcore.png}
%{Xrays.pdf}
\vspace{-0cm}
\caption {\small {\bf Left}. The central 60pc of 30~Doradus in H$\alpha$ using a public domain image taken with the 2.2m/WFI imager on La Silla. North is on top; East is on the left. The  {\bf Right}. Chandra 20-year anniversary X-ray image of the same region. The two images have been astrometrically matched using point-like X-ray sources and the software package Aladin. The cross near the center shows R136, the core of the ionizing cluster. (the second cross on the X-ray image shows one of the astrometric references). The red and blue dots show the positions of red and blue shifted single profiles relative to the systemic velocity of 30Dor, $V_{30Dor}=265.4$~\kms. The orange crosses mark positions where the profiles are double and the green dots show the positions of the multiple (mostly triple) profiles.The yellow line draws the contours of the giant molecular cloud 30Dor-10 \citep{inde13}.}
\label{fig110}
\end{figure*}

{ Single profiles delineate bright (dense) \halfa filaments, and lie at the edges of bright X-ray zones. These dense filaments define the "head" of the Tarantula, which, as shown by the distribution of blue-shifted and red-shifted profiles (blue and red dots), consists of several expanding bubbles that are the mouth and the eyes of the spider.  The large elliptical bubble that surrounds R136 is asymmetrical with the eastern side being more luminous in X-rays. This "mouth of the Tarantula" opens towards the east indicating that the hot gas is escaping through the mouth to inflate the giant bubble Chu-Kennicutt\#2, as one would inflate a balloon for a birthday party!

There is another expanding bubble due North of R136 that is very bright in X-rays and that is open to the north toward the giant Chu-Kennicutt\#5 shell. Similarly the bubble due south of R136 seems to be inflating Chu-Kennicutt\#1.  So we get the impression that the dense filaments that define the head of the Tarantula act as nozzles that channel the cluster wind from the source (R136)  out to the giant Chu-Kennicutt shells.

The clumpy structure of the hot cluster wind seen in the X-ray image is mostly due to shadowing by the dense dust+gas structures that can be clearly appreciated in the \halfa image. The X-rays are faint in places where the \halfa emission is bright and blue-shifted, indicating that the nebular component is in front of the hot gas, and viceversa. Receding filaments seem to lie behind that hot gas, but not always; the bright loop directly NW of R136, where the X-rays are faint, is redshifted. However, the giant shell Chu-Kennicutt\#3 is adjacent so it is possible that the hot wind has already escaped from that region.

\subsection{Kinematics}
 
To compare with the results of \cite{chuken}, Table~\ref{kine} presents an estimation of the kinetic energy budget of the nebular gas. The total kinetic energy is $2K_{tot}=M_{gas}\sigma_0^2$ where $\sigma_0$ is the integrated velocity dispersion of the gas estimated from the single-Gaussian fit of figure~\ref{fig0} and $M_{gas}=4\times10^5$\msun\  is the total mass of nebular gas according to \cite{chuken}. The values for the kinematical core and halo are obtained assuming that the mass of gas is proportional to the \halfa luminosity.

The kinetic energy of the "general" turbulence was calculated as follows. We saw that the single \halfa profiles as well as each of the two components of double \halfa profiles have average widths of $\sim11.8$~\kms. We do not know the source of energy that sustains these motions, but we do know that when we observe the same profiles in the forbidden lines, they are resolved into many narrower components. We also ignore the mass of these turbulent shells and filaments, but, since they comprise the brightest regions of 30Dor, it is reasonable to assume that they contain most of the nebular mass (assuming that the mass is proportional to the \halfa luminosity) . We calculated the velocity dispersion in two ways: directly from the average profile widths (min) and convolving the profile width with the radial-velocity dispersion of the single profiles (max),  $\sigma_{max}=\sqrt{11.8^2+11.6^2}=16.5$~\kms.}

\begin{table}[!h]
\vspace{0cm} 
\caption{\bf Kinetic energy budget}
\tabcolsep 2mm
\small
\begin{tabular}{ l c c c }
\hline\hline
Component       		& $<\sigma>$   &  Mass  			& $K/K_{tot}$      	\\ 
                          		& \kms 		& (\% total)		& 				\\ \hline 
Kinematical Core 		& 23.7 		& 23				& 0.24			\\ 
Kinematical Halo  		& 23.5		& 77    			& 0.76			\\  
Turbulence min.	       	& $11.8$		& 90			      	& 0.25			\\  
Turbulence max.		& $16.5$		& 90				& 0.45			\\  \hline
%Double profiles shells	&			&				& 0.99			\\  \hline
\end{tabular}
%\tablefoot{
%\tablefoottext{b}{ \small From Fig.\ref{fig6};}
%\tablefoottext{b}{ \small }
%}
\label{kine}
\end{table}

Since at least some, and probably much, of the radial-velocity dispersion of the single profiles is due to expansion motions, the velocity dispersion of 16.5~\kms\ given in the table is a maximum. Thus, the fraction of the total kinetic energy contained by general turbulence is closer to 25\% than to 45\%, compared to 50\% estimated by \cite{chuken}.

\section{Turbulence}

\subsection{Kolmogorov turbulence}

Observations of the other prototypical GHR in the local group, NGC604, by  \cite{medina} appear to show a well developed Kolmogorov-like kinetic energy cascade in that giant nebula. They found that the structure function,
 \begin{equation}
S_2({\bf r}) =<[v({\bf x})-v({\bf x}+{\bf r})]^2>
\end{equation}
-- the square of the difference in radial velocity between two points separated by a distance $r$ averaged over all pairs of positions separated by $r$ --  is a power-law of slope $\sim5/3$ for $r\le10$pc and flattens at larger separations. 

Since by definition $\sigma\equiv\langle v^2\rangle-\langle v\rangle^2$, if $\langle v\rangle=0$ the structure function $S_2=2\sigma^2$, so it is convenient to normalize by the square of the the radial-velocity dispersion $\sigma_{rv}$ such that the normalized function peaks at $S_2\sim2$  \citep{arthur2}. Figure~\ref{fig9} reproduces the normalized structure function of NGC604  calculated using the same radial velocity data of  \cite{medina} kindly provided by Gustavo Medina-Tanco.

\begin{figure}[h!]
\vspace{3.8cm}
 \includegraphics[height=4.1cm, trim= 0.5cm 0 8cm 20cm]{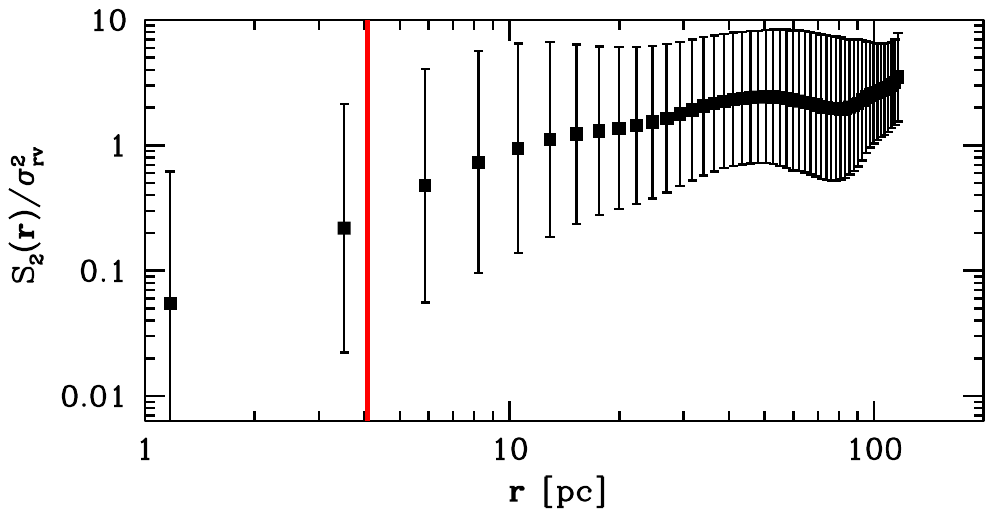} 
\vspace{-3.3cm}
\caption{Structure function of the central region of NGC~604 using the same radial velocity data from \cite{medina}. The red bar shows the radius (4.1pc) corresponding to the seeing disk ($1''$) of the TAURUS-II observations used to construct the structure function.}
\label{fig9}
\end{figure}

%The authors say: "In 30Dor the profiles at large radii tend to be single and have small radial velocities as expected in a virialized system where the orbits are predominantly radial and at large projected radial distances have small velocities perpendicular to the plane of the sky." but the word "orbits" biases the reader with the analogy to ballistic objects (probably the authors' bias too) when it is only "motion" of some type that is radial. The word "motion" would be better than "orbit" as the authors have not shown that the gas particles move around in looping paths inside the nebulae. OK!!!

We used our full sample to construct the structure function for 30Dor as for NGC604, but using three different ways of measuring the radial velocities. The top panel of Fig.\ref{fig9a} corresponds to the method adopted by \cite{medina} of fitting single Gaussians to all profiles. The second, shown in the middle panel, uses the radial velocity of the strongest peak at each position; and the third uses only the single profiles. Typically each point in the top two panels is the average of 16000 to 78000 pairs, except for the smallest separation that has only 6500 pairs.  The vertical bars represent the rms dispersion around each average. 

\begin{figure}[h]
\vspace{4.5cm}
\includegraphics[height=8cm, trim= 0.5cm 0 8cm 12cm]{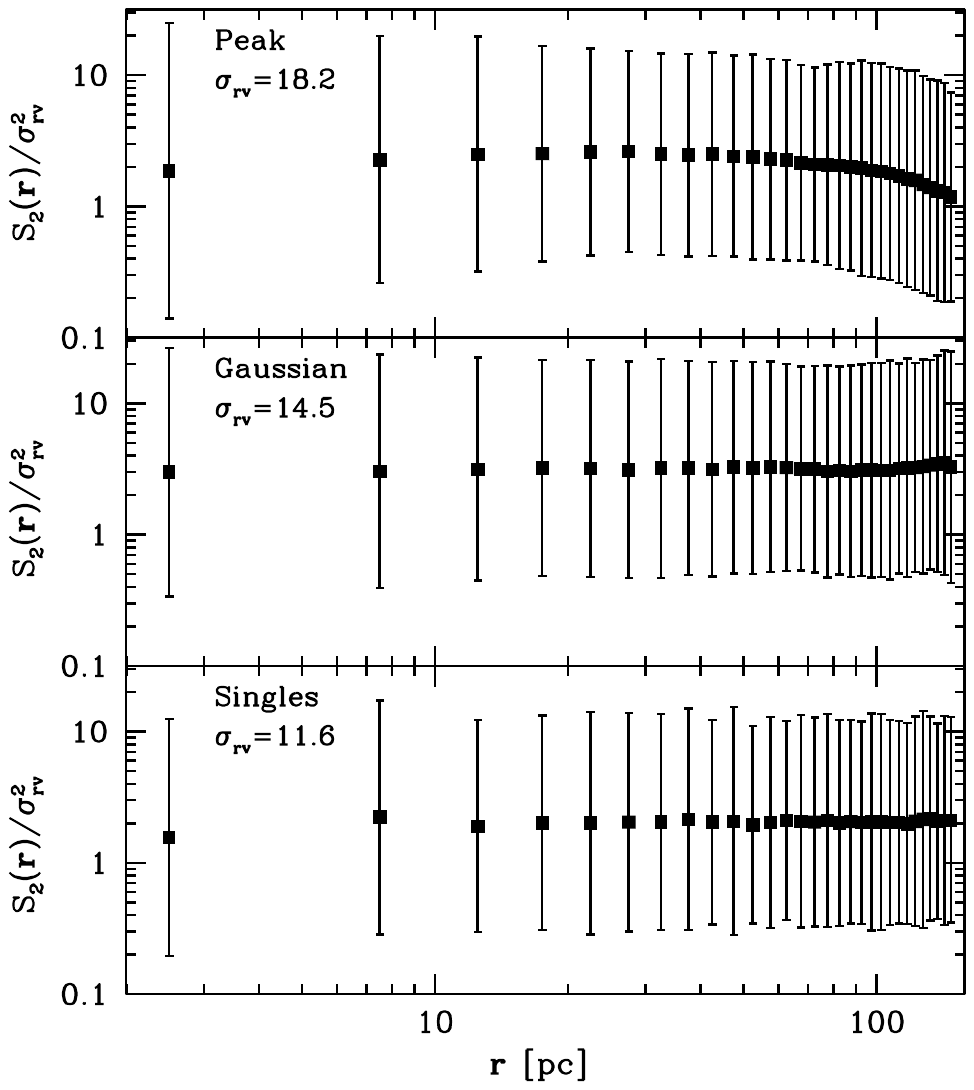}\vspace{-0.3cm}
\vspace{-2.7cm}
\caption {The normalized structure function of 30 Doradus. {\bf Top}. Using radial velocities measured from single-Gaussian fits; {\bf Mid}. Radial velocities of the strongest peak of each profile; {\bf Bottom}. Using only profiles classified as single.}
\label{fig9a}
\end{figure}

Contrary to that of NGC604, the structure function of 30Dor has no structure: it is flat!  Since the $L-\sigma$ relation of Fig.~\ref{lsigma} implies that the kinematics of GHRs depends only on their luminosities, we expected 30Dor and NGC604 to have similar, if not identical, structure functions, which turned out not the case. Why?   

Unfortunately, there are several problems and inconsistencies with the results of \cite{medina}. The first and most obvious is seeing. The red bar in Fig.~\ref{fig9} at 4.1pc corresponds to the seeing ($1''$) reported for the Taurus-II observations of \cite{saba} upon which the structure function is based.  The separation of the first two points in the structure function is less than the seeing, while the next two points at $<2''$ are also affected by seeing. These four points define the slope of the function for $r<10$pc.}
Removing these four seeing affected points the remainder of the structure function of NGC604 is basically flat.

A second issue that may blurr the comparison between 30Dor and NGC604 is beam-smearing. At the distance of M33, $1''=4.1$pc,, while the data of \cite{medina} were binned to $2''\times2''$, which results in a significant smoothing of the radial velocities. There are also some problems with the TAURUS-II data for NGC604.  For example, the TAURUS-II profiles of \cite{casiana} are substantially broader than the TAURUS-I (plus Echelle) results of \cite{chuhui}.  We were unable to recover the original data to check these issues, so for the moment it seems safer to ignore the results of \cite{medina}, although we cannot overemphasize the importance of re-observing NGC604 hopefully at a higher spatial resolution.

\subsection{Gravity}
{ \cite{chuken} dismissed gravity as an important source of kinetic energy because "...even the largest imaginable mass for 30Dor would be far too small to explain the velocity dispersion of the gas inferred from the width of the integrated profile", $\sigma_0=23.5$~\kms (Fig.~\ref{fig108a}). But we know, as they knew, that a major part of this width is furnished by expanding shells. So let us revisit the issue.

We calculated the structure function of an N-body model of a $10^3$\msun\  virialized cluster containing  2000 particles of equal mass (kindly provided by Sverre Aarseth) shown in Figure~\ref{fig9b}, 

\begin{figure}[h]
\vspace{-.8cm}
\includegraphics[height=8.5cm, trim=1.3cm 0cm 5cm 12cm]{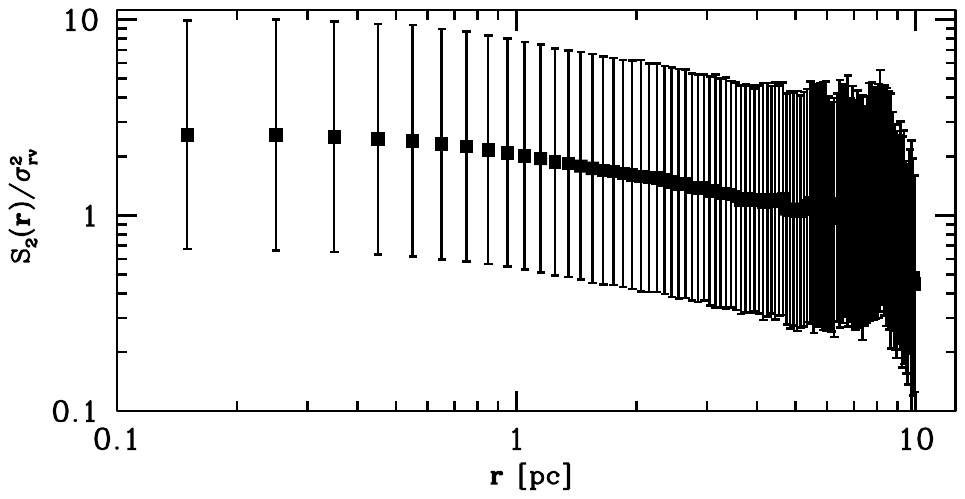}
\vspace{-3.3cm}
\caption {Structure function of a virialized cluster of 2000 particles. All particles have equal masses and the total
mass of the cluster is $10^3M_{\odot}$.  For a virialized system $S_2({\bf r})\simeq2\sigma^2$ and the velocity dispersion $\sigma$ is constant in the core.}
\label{fig9b}
\end{figure}

The match is not perfect, but still provides a much better match than a Kolmogorov turbulent energy cascade. This does not prove that gravity is driving the turbulence in 30Dor, but at least provides a consistency check. 

Another way of probing the effect of gravity is to estimate the mass required to generate the observed motions through the virial theorem. The morphology of 30Dor and the results of \cite{chuken} show that the outer parts of the nebula are dominated by expanding motions, so here we will restrict our analysis to the kinematical core. 

Figure~\ref{solima} shows the radial-velocity dispersion $\sigma_{rv}$  (in red) and the mean velocity width (in blue) of the single profiles plotted as a function of radius. Each bin contains 50-70 points except for the first bin ($10<r\leq25$pc) that contains 98 single profiles. We excluded the central 10pc that is completely dominated by expansion motions. 

\begin{figure}[h]
\vspace{3cm}
\includegraphics[height=7.5cm, trim= 2.0cm 0 8cm 10cm]{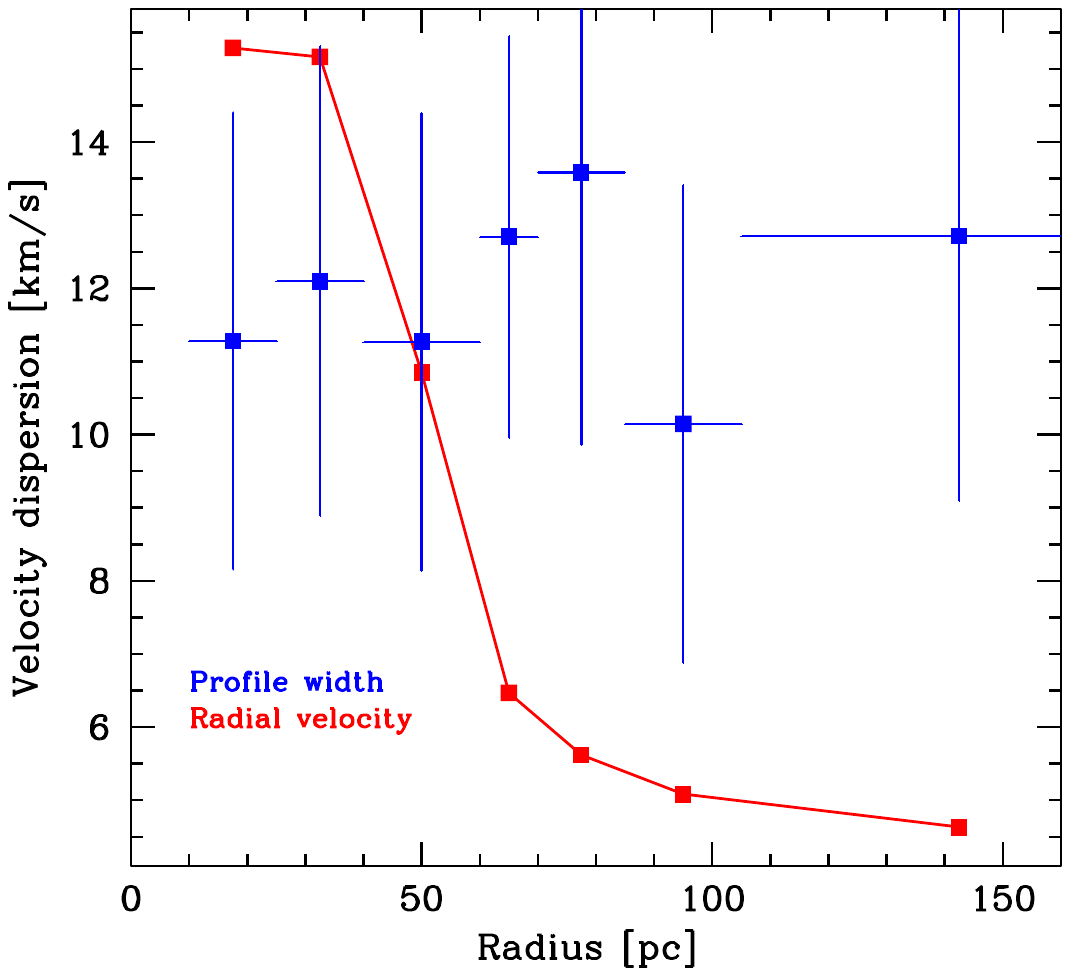}
\vspace{-3.0cm}
\caption {Velocity dispersion as a function of radius from the single profiles. The red points and line show the radial-velocity dispersion of the profile centroids and the blue points show the average width of the profiles in each radial bin. The horizontal error bars show the size of the bins, and the vertical bars show the rms dispersions of the average widths.}
\label{solima}
\end{figure}

The  average width of the profiles in the kinematical core is $<\sigma_{core}>=11.2\pm3.0$~\kms\ and the radial-velocity dispersion is $\sigma_{rv}=15.3$~\kms. The respective minimal and maximal turbulent velocities in the core, therefore, are $\sigma_{min}=11.2$~\kms\ and $\sigma_{max}=\sqrt{11.2^2+15.3^2}= 19$~\kms. Assuming an effective radius of 10pc (instead of the 100pc used by \citealt{chuken}), the corresponding mass limits from the virial theorem are,

$$ 2.9\times10^5 \leq M/M_{\odot} \leq 8.4\times10^5  \eqno (2) $$

The observed (photometric) mass of the kinematical core is quite uncertain, the stellar mass.being probably the least uncertain. \cite{selman1999} quote a mass of $M_*=68000$\msun\ that results from integrating the observed (power-law) IMF between 3\msun\ and 120\msun. The low-mass (m<3\msun) IMF of 30Dor is not known, but the IMF of most young star forming regions turns over at masses $\sim0.5$\msun\  \citep{bastian}. Extrapolating the IMF of  \cite{selman1999}  down to 0.5\msun, yields a total stellar mass of  $M_{\star}=1.5\times10^5$\msun\ for 30Dor, 50\% larger than the value generally quoted in the literature (e.g. \citealt{doran} and references therein), 

The mass of ionized gas is more uncertain.  Here we will adopt the value of $M_{HII}=4.5\times10^5$\msun\  from \cite{chuken}, which according to these authors is uncertain by factors of 2-4.  \cite{ox} derive a similar mass with an uncertainty of "only" a factor of 2 (0.2dex).

The most uncertain component is the mass of neutral and molecular gas also discussed by \cite{ox}.  A comprehensive review of this contribution is well beyond the scope of the present paper, but in Appendix~\ref{ap2} we present a summary and justify our optimistic value of $M_{HI+H_2}=9\times10^4$\msun\  for the amount of neutral and molecular gas within the nebular core.

The core of 30Dor contains the bulk of the stars and 23\% of the \halfa luminosity, so the total mass within the kinematical core is,
$$ M_{core}=M_{\star}+0.23\times M_{HII}+M_{HI+H_2} = 3.4\times10^5~M_{\odot} \eqno (3) $$
which is comfortably above the minimum value required by Eq.2. So, again, we may have consistency, but not proof. 

%Since 30Dor-10 and the surrounding dense \halfa filaments have similar blue-shifted radial velocities, it is reasonable to conclude that the molecular cloud is being blown away from the central region of 30Dor by the radiation pressure. Under this assumption, we can obtain a rough estimate of the central mass of 30Dor by simply balancing the gravitational pull on a cloud of $1.8\times10^5$\msun\  \citep{chevance} with the radiation pressure at 30pc \citep{lopez} on an area of $10\times10pc^2$. This gives a mass of $2\times10^5$\msun, and an escape velocity of $\sim7$~\kms much lower than the mean radial velocity velocity of the 30Dor-10, $\Delta v=-16~$\kms\ \citep{inde13}. The value of eq.2, therefore, is a conservative upper limit.
 
A simple and direct proof would be to show that the velocity dispersion of the stars is consistent with that of the gas. Unfortunately, however, this test turns out to be neither simple nor direct. \cite{Bosch01} and \cite{Bosch09} found a surprisingly large velocity dispersion for NGC2070 that can  be explained if a large fraction of the massive stars in the cluster are binaries. Subsequently, binary orbits have been measured for many of these stars confirming that conclusion \citep{jenot12a}. There are several other complications related to mass segregation and rotation \citep{jenot12b} that compromise the putative simplicity of the test, but the observations show that, at least in the cluster core ($r<5$pc), the stars are in virial equilibrium  \citep{jenot12a, jenot12b}.

To conclude this discussion, let us revisit the mass required to explain the integrated velocity dispersion by gravity. Adding up the HII mass ($4.5\times10^5$), the stellar mass ($1.5\times10^5$), and the molecular mass ($1.8\times10^5$) the total mass of the nebula is $M_{30Dor}=7.8\times10^5M_{\odot}$. The half-mass radius, therefore, is close to the radius of the kinematical core ($r=25pc$), but much smaller than the value of 100pc assumed by \cite{chuken}. The viral mass for a velocity dispersion of 23.5~\kms\ and an effective radius of 25pc is $3.2\times10^6M_{\odot}$, still significantly larger than our optimistic value for $M_{30Dor}$, but by no means "the largest imaginable"! 

\subsection{Small scale turbulence}

We have seen that there is no large scale supersonic hydrodynamical turbulence in 30~Doradus, and that the gravitational potential is not strong enough to furnish all the kinetic energy of the gas: The large-scale structure of 30Dor is organized by the cluster wind, but the wind-driven shells are themselves turbulent. 

This turbulence could be due to Kolmogorov-like kinetic energy cascades powered by the wind, such as observed in the Orion Nebula \citep{arthur2}. However,
even within the shells and filaments of 30Dor, we are looking spatial scales orders of magnitude larger than the turbulent regions in Orion, so we have no hope of resolving the turbulent cascade. However, our data (mostly HARPS but also some FLAMES profiles) show that the forbidden profiles can be resolved into many components with radial-velocity dispersions similar to the widths of the profiles. Such mutiplicity would explain why at the same positions the \halfa profiles are subsonic  but highly supersonic in the forbidden lines.
 
An intriguing possibility to explain the small scale turbulence is that sections of the leading shocks composing the large structures loose energy by radiation as the medium they sweep piles-up in front, so necessarily secondary shocks emanate from the hot gas to transmit the interior pressure. These "shockletts" slam first onto swept-up gas and eventually onto the leading shocks giving them small jolts that here we relate to the structure we observe in the forbidden lines.
 
The ionization of the 30Dor nebula is by and large provided by the massive stars, with a small contribution from fast shocks that, in principle, should be more important in the denser parts of the nebula \citep{pelegrini}. Figure~\ref{shark} shows the shock sensitive diagnostic diagram that can be obtained using the FLAMES spectra, color coded according to the distance to R136. 

 \begin{figure}[h]
\vspace{3.0cm}
\includegraphics[height=7.7cm, trim=0.8cm 0cm 5cm 12cm]{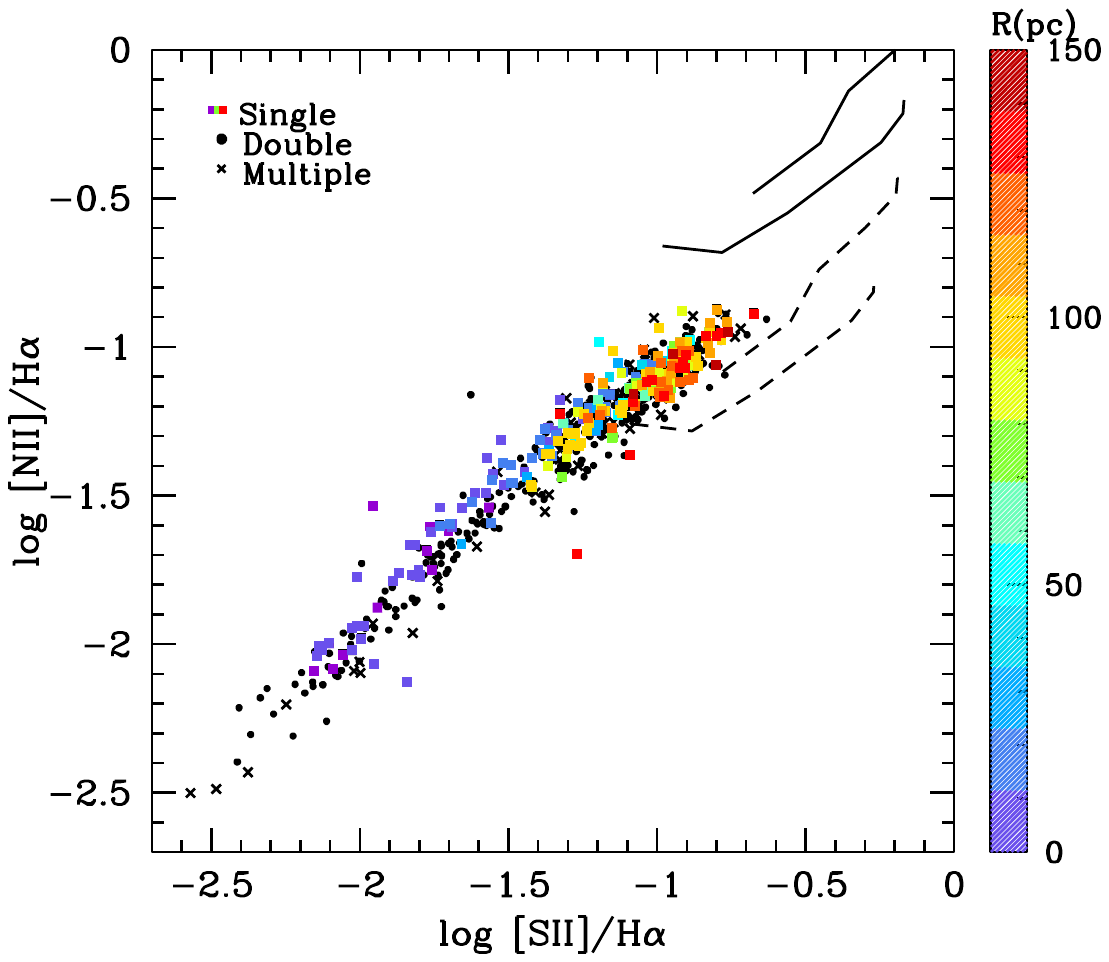}
\vspace{-2.8cm}
\caption{Point-to-point nebular diagnostic diagram for the different types of profiles described in the text. The points for the single profiles are colored according to the distance to R136 shown by the scale on the right. The solid curve shows the low-velocity shock models of \cite{daputa} for a range of velocities between 150~\kms and 500~\kms; two values of the magnetic field density (0 and 1 $\rm \mu Gcm^{-3}$); and solar abundances. The dashed curve shows the same models scaled to the metallicity of 30~Dor from \cite{pelegrini}.}
\label{shark}
\end{figure}
The figure shows that emission from low velocity shocks is clearly visible in the low-density external regions of the nebula, but not in the densest regions in the core of the nebula. This does not mean that shocks are not present in this region; it only means that, if present, they are completely overshadowed by photoionization at optical wavelenghts,  reflecting, perhaps, how the cooling process operates behind the leading shocks as these continue to ram into a cloudy medium of very different densities. This also determines the speed of the shockletts driven into the nebular swept-up gas as this cools and approaches the nebular equilibrium temperature. 
{\subsection{The broad unresolved component}

Our analysis thus far has ignored the mysterious broad unresolved component that pervades the nebula and that is seen both in the Balmer lines and in the forbidden lines of [NII], [SII], and [OIII], precisely because we ignore its origin.  We have seen, however,  that like the narrow components, the broad unresolved component of the forbidden lines is narrower than H$\alpha$, so it seems reasonable to conjecture that the physical conditions of the gas that emits the broad unresolved lines are similar to those of the nebula as a whole, and therefore, that it is probably due to a large collection of fragments of shells and filaments left behind by the cluster-wind as it successively breaks out of shells of swept-up material. 

We have not explicitly included this component in our estimates of the kinetic energy budget of the nebular gas, for which we used single-Gaussian fits. Although the width of the broad component is large, its residual \halfa flux relative to the single Gaussian fit is negligible, so we can safely ignore the contribution of the broad component to the kinetic energy budget. 

\section{Discussion}

There is broad disagreement in the literature about the structure of 30Dor and its genesis  \citep{pelegrini,lopez}. Below we present a summary of our own view, which is the result of a series of in-depth investigations about the formation and evolution of super-star clusters and their associated Giant HII Regions  (\citealt{silich} and references therein), and which we will use in the interpretation of the observations presented in this paper.

\subsection{The genesis of Giant HII Regions}

The progenitors of GHR are Giant Molecular Clouds (GMCs) that are clumpy and highly turbulent as a consequence of multiple prior generations of star formation, the gravitational field of the parent galaxies, and their own gravitational potential, which is substantial given their large masses (e.g. \citealt{heyerdame, krumholz}). Even before any stars form, the turbulence within these GMCs is supersonic (with respect to the sound speed of the molecular gas  ($\sim1$~\kms), causing the condensations to accrete more molecular material as it cools behind their leading shocks. This sets the initial conditions at the onset of star formation that occurs within the largest and most massive condensations.

The arrival of the most massive stars on the main sequence (MS) produces a sudden ample supply of UV photons that changes the sound speed to $\sim10$~\kms\  as the gas becomes photoionised.  At the same time, the strong stellar winds from these stars (that reach the MS first) collide with neighboring winds generating multiple shocks that thermalize the winds causing a large overpressure in the star forming volume that ends up driving the shocked stellar wind gas as an isotropic supersonic {\em cluster wind } (v$_{cw}\sim1000$~\kms ) into the surrounding medium \citep{gttelles}.

This cluster wind immediately encounters a large number of neighboring condensations, some of them fully or partially ionised, and the interaction leads to another (global) reverse shock that heats up the incoming gas to very high temperatures ($T\sim10^7$K). The resulting hot gas rapidly expands along multiple paths of least resistance and flows away into lower pressure regions engulfing the dense ionized condensations that were originally part of the parent GMC. The expansion of this continuous wind generates secondary shocks that sweep-up the low density medium between condensations producing large bubbles and super-bubbles that eventually break-up allowing the hot gas to escape and create the hierarchy of shells, tubes, and filaments that characterize GHR. The expansion eventually reaches the edge of the GMC and the cluster wind escapes into the intergalactic medium carrying with it the products of stellar evolution \citep{gttelles}.

\subsection{The genesis of 30 Doradus}

The morphology of the gas (nebular and X-rays) confirms that the structure of 30~Doradus was carved by the cluster wind and not by the winds of individual stars; the coherent \halfa structures that we see are seldom, if ever, centered on individual stars. 

We have shown that the clumpy structure of the X-ray emission is due to shadowing by dense clouds of dust and nebular gas, and also by the giant molecular cloud located NNE of R136, which we claim is in front of the nebula on the basis of the radial velocities of all its individual clumps.

The innermost dense regions of the nebular gas have been pushed into coherent structures by the cluster wind in such way that they form nozzles through which the hot gas flows towards the outer regions of the nebula thus creating the giant expanding shells that characterize 30~Doradus.  The Chandra X-ray images of \cite{townsley} show regions - notably the large Northern plume - where the cluster wind is already escaping from the nebula. The X-ray gas occupies regions of low \halfa surface brightness where the profiles are inevitably double, clearly showing that the expanding shells are being "inflated" by the hot cluster wind. 

30~Doradus has a clear core-halo structure evidenced mainly by the brightness. However, the overall kinematics of the core and the halo are surprisingly similar; the integrated profiles are virtually identical and only differ in the broad wings albeit slightly. This may just be a coincidence given that the way the cluster winds carve the original molecular cloud must depend on the original configuration, which certainly must vary from one nebula to the next. It would be interesting, however, to verify whether other GHR in the Local Group present similar "coincidences". 

The remnants of the dense clumps of the placental GMC have been piled together by the cluster wind to form the densest parts of the nebular gas, mostly in the form of long twisted filaments. At small scales, the forbidden line-profiles of these shells and filaments are supersonic and composed of many narrower components, which could be due to the detailed hydrodynamics of the expanding shells.

The source of energy for the whole process, therefore, is the ionizing cluster: the kinematics through the cluster winds and the emission-line luminosity through the ionizing radiation. Ultimately, this explains why $L$ and $\sigma$ are so tightly related, although it remains hard to explain the steep slope of the $L-\sigma$ relation.  The relation implies that the velocity dispersion of the gas increases very slowly with the mass of the ionizing cluster ($\sigma\sim M_{cl}^{1/5}$), so explaining the $L-\sigma$ relation remains a challenging problem in astrophysics. Pure gravity would have been much simpler!

\section{Summary and conclusions}

Theoretical studies predict that the formation and evolution of Giant HII regions is determined by hot cluster winds that result from the merging and thermalization of the winds from the individual ionizing stars. Our observations largely support this scenario: none of the large nebular structures that characterize the nebula, with the exception of the central bubble that surrounds R136 are centered around individual massive stars. The hierarchical structure of the shells, on the other hand, is exactly what is expected from a cluster wind that originates in the core of the ionizing cluster (R136) and breaks out of successive shells built by the material swept up by the same wind.           

We have confirmed previous results that the kinetic energy of the nebular gas is mostly in the form of the expansion of the multiple shells. Not only are the \halfa profiles in the low surface brightness regions inside these shells double or multiple, but also the profiles of the rims of these shells, which are mostly single, clearly show organized expansion motions.  Expansion accounts for $\sim70$\% of the kinetic energy while the remainder is manifested by the supersonic intrinsic widths of the individual profiles.
%The gravitational potential of stars and gas is sufficient to provide the kinetic energy required by the profile widths, which therefore, can be interpreted as the residual orbital  motions of the, now ionized, dense clumps of the original molecular cloud that spawned the 30~Doradus superstarburst cluster. 

The radial dependence of the kinematics and of the \halfa surface brightness clearly show a well defined core-halo structure.  The nebular core ($R\leq25$pc) contains the bulk of the ionizing stars and 23\% of the total \halfa luminosity. The massive cluster wind that carves the nebular structures originates in the core and is channeled to the external giant expanding bubbles through break-out structures within the core, which act as nozzles.

The clumpiness of the (hot thermalized cluster wind) X-ray emitting gas is due to shadowing by foreground nebular gas and dust, so it is reasonable to conclude that the hot gas uniformly fills the nebular core, and is expanding through the break-out nozzles to inflate the characteristic giant (Chu-Kennicutt) external bubbles of 30~Doradus.

The structure function of the turbulence of the nebular gas in 30Dor is flat. The idea that the supersonic turbulence of the gas in Giant HII Regions is due to a turbulent Kolmogorov-like kinetic energy cascade put forward by \cite{medina} can be definitively ruled out. Instead, we find that the structure function of a virialzed cluster is more compatible with our observations, albeit only within the cluster core where the velocity dispersion is constant.
 
The virial mass corresponding to the observed global velocity dispersion of $\sigma_{30Dor}=23.5$~\kms\ and an effective radius of $R_{eff}=25$pc is $M_{vir}=3.2\times10^6$~\msun. The main source of the supersonic velocities observed in the nebular gas is the hot cluster wind, with a small but significant contribution from the gravitational potential of ionized gas, molecular gas, and stars. The total (photometric) mass of 30~Doradus is a few times $10^5$\msun\ with large uncertainties but certainly less than $10^6$\msun.  
 
%Conclusion #1 is completely unproven in this paper, as documented above. The connection with NGC 604 is vague throughout this paper, which is mostly about 30 Dor.

%conclusion #3 seems like speculation. The authors do not present evidence that helps clarify what the broad components are.

%Based on our analysis of the kinematics of the nebular gas in 30~Doradus and the comparison between the morphology of the gas and dust with the X-ray emission map we conjecture that  the nebula may be confined by a huge bubble of gas and dust.  Thus, the hot X-ray emitting plasma has already expanded to fill the entire nebula such that the dense shells seen in the nebular and dust emission mask the hot plasma, but do not contain it.

%In a forthcoming paper we intend to return to some of the issues raised in this paper by mapping the density and ionization structure of 30~Doradus. This will allow us, in particular, to estimate the mass of the nebular gas and to study the energy balance between the expanding shells and the power supplied by the stellar winds.
 %\section{Concluding Remarks}
 
 \section{Acknowledgements}
Part of this work was done during the 24th Guillermo Haro Workshop at INAOE We are grateful to the organizers of the meeting and to INAOE for creating and maintaining this unique venue. 
We thank  Sverre Aarseth for providing his n-body simulation of a virialized cluster 
and Gustavo Medina-Tanco for sharing with us the Taurus-II radial velocities of NGC~604. We also thank the PI's of the observing programs that spawned the treasure trove of data that we have used in this paper, and ESO for making these data reduced and readily available through its Phase3 archive. In particular, we thank Reinhard Hanuschik for helping us understand some subtleties of the pipe-line reductions of FLAMES/GIRAFFE data. JM acknowledges the hospitality of ON and INAOE where parts of this paper were written.

\bibliographystyle{aa}  
\bibliography{bibi}  
%\newpage 
\appendix\rm
\section{Geometrical effects related to radial gradients in the strength of the forbidden lines}
\label{ap1}
The strength of the forbidden lines is inversely proportional to the electron density $N_e$, which decreases with radius roughly as $N_e\propto R^{-0.7}$ as shown in Figure~\ref{app1}.

\begin{figure}[h]
\vspace{2.7cm}
\includegraphics[height=6cm, trim= 0.5cm 0 8cm 12cm]{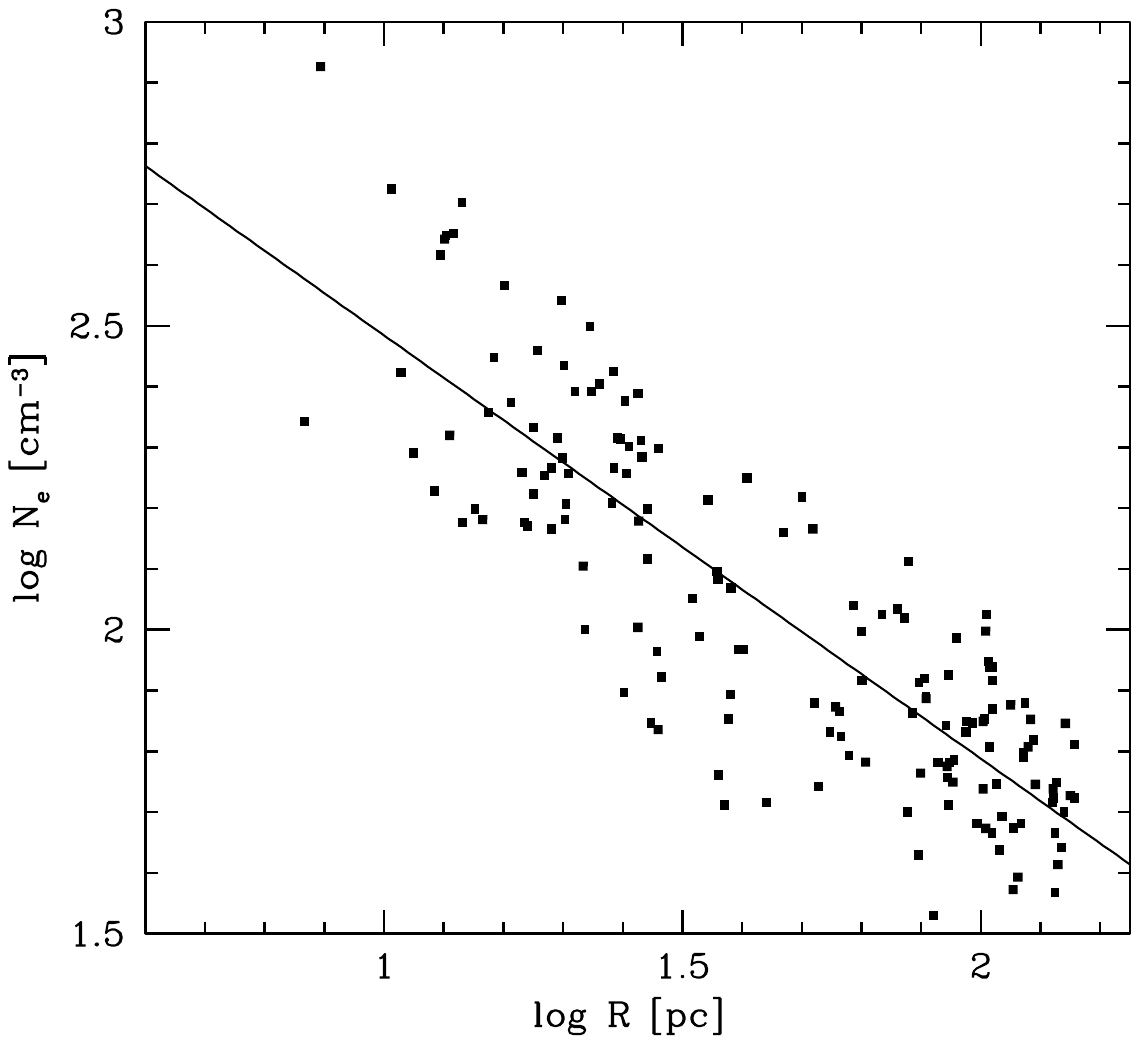}
\vspace{-2.cm}
\caption {Logarithmic plot of the electron density derived from the [SII] doublet as a function of radius. The line shows a least-squares fit of slope -0.7}
\label{app1}
\end{figure}

The flux of ionizing radiation should decrease roughly as $N_{\nu}\propto R^{-2}$ since a large fraction of the ionizing stars is concentrated at the center of the ionizing cluster (R136). However, not all ionizing stars are in R136 \citep{castro}, so it is safer to assume that $N_{\nu}\propto R^{-\alpha}$ where $\alpha$ measures the central concentration of the ionizing flux. Thus, the ratio of [NII]/H$\alpha$ scales as $R^{\alpha-0.7}$.

Figure~\ref{app2} plots this ratio as a function of radius for the single profiles. The least-squares fit shows that $F([NII]/F(H\alpha) \sim R^{0.2}$ in the halo and $\sim R^{0.7}$ in the nebular core (we used only [NII] that has more points). This indicates that $\alpha=1.4$ in the core and $\alpha=0.9$ in the halo. As an exercise, keen readers can compare these {\em predictions} with the observed distribution of massive stars using the catalog of 30Dor stars \citep{castro}. 

\begin{figure}[h]
\vspace{2.7cm}
\includegraphics[height=6cm, trim= 0.5cm 0 8cm 12cm]{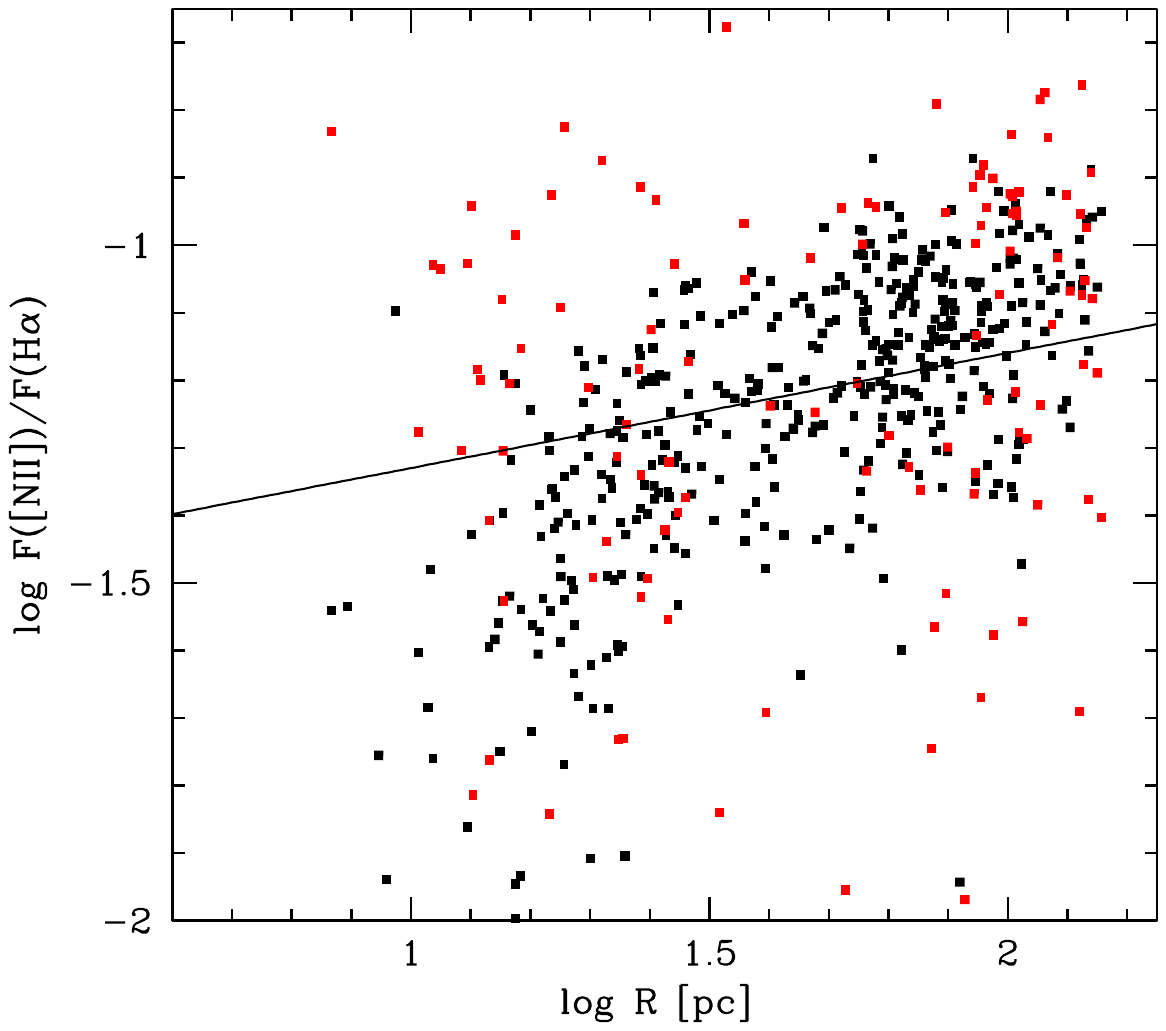}
\vspace{-2.cm}
\caption {The black points show the ratio of the flux of [NII]678.3 to H$\alpha$ as a function of radius. The red points correspond to [SII]/\halfa. }
\label{app2}
\end{figure}
 
In principle, therefore, one could expect the difference between the widths of [NII] and H$\alpha$ to depend on radius, but this is not the case. Figure~\ref{app3} plots the difference in the widths of [NII] and H$\alpha$ for the single lines in our sample. We observe no obvious trend. We exclude, therefore, that the H$\alpha$ lines are broader due to geometrical effects where [NII] (and [SII]) probe mostly the kinematics of nebular halo whereas H$\alpha$ probes the entire nebula.
 \begin{figure}[h]
\vspace{2.9cm}
\includegraphics[height=6cm, trim= 0.5cm 0 8cm 12cm]{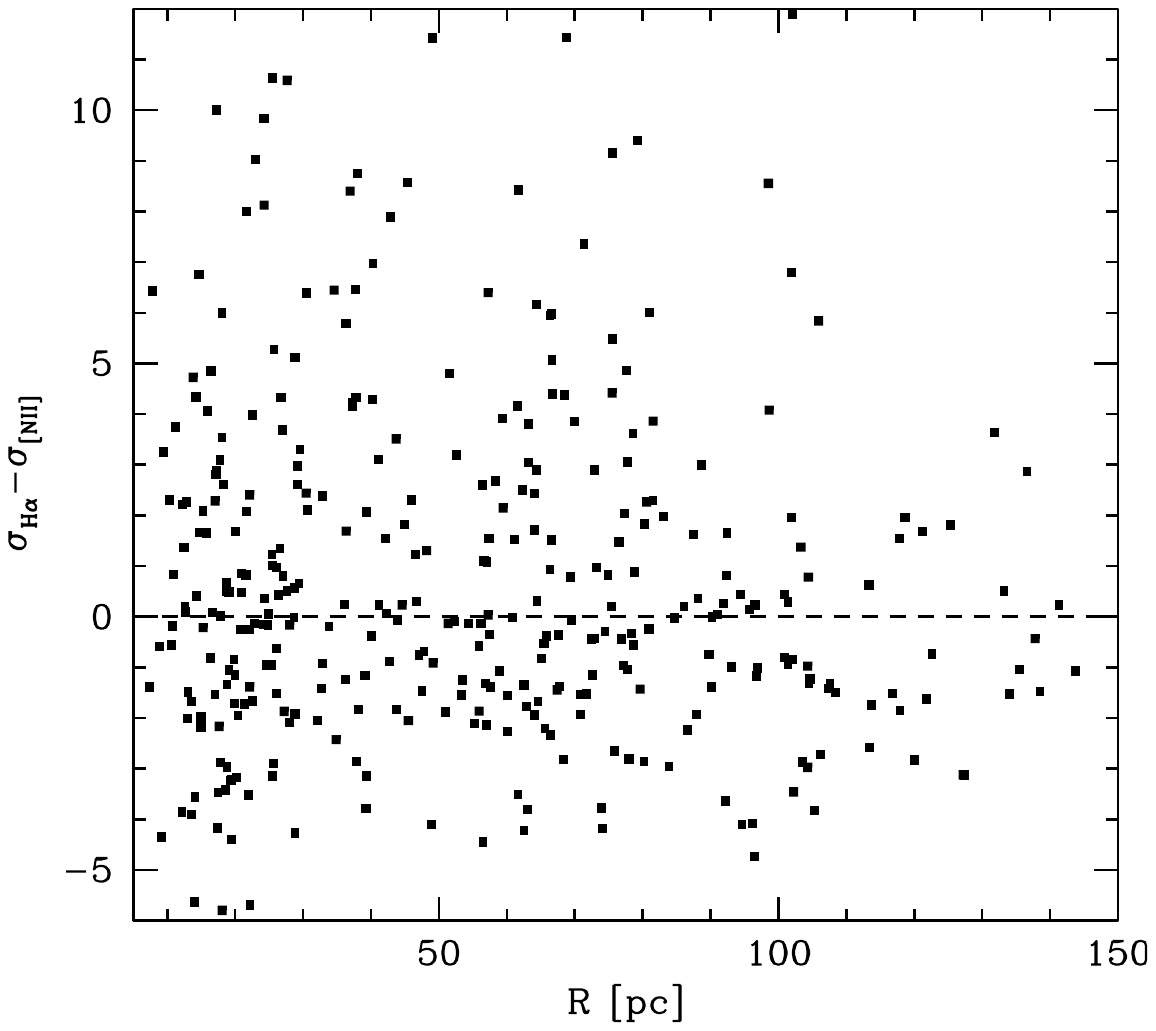}
\vspace{-2.cm}
\caption {Difference  between the rms widths of the individual [NII] and H$\alpha$ profiles.}
\label{app3}
\end{figure}

\section{Neutral and molecular gas in 30 Doradus}
\label{ap2}

We refer to the work of \cite{ox} for a recent review of the gas content of 30Dor.  The HI mass estimated by \cite{ox} using the 21cm data of \cite{kim99} increases as $M\sim R^2$ as expected for a uniform constant surface density sheet of HI extending out to at least 250pc from R136. On the other hand, the profiles of $H_2$ and HII reach a plateau at $R\sim30$pc). Unfortunately, the list of 21cm LMC sources of \cite{kim99} inexplicably misses GS78, which corresponds to 30Dor, so we are not able to check whether the HI-sheet has the same radial velocity as the \halfa emission. Therefore, we have taken the view that 30Dor does not contain significant neutral hydrogen, certainly not in the core.

There is, however, a giant molecular cloud - 30Dor-10 - projected at a distance of 15pc-30pc north of R136 (illustrated by the yellow contour in Fig.~\ref{fig110}.  ALMA high resolution observations \citep{inde13} give a total mass of $M_{H_2}=5.5\times10^4$\msun\ for 30Dor-10, but \cite{chevance} argue that most of the molecular gas in that region may be "CO-dark".  Using a IR observations and a PDR model, the find derive an $H_2$ mass of $1.8\times10^5$\msun.  

\begin{figure*}
\hspace{0.cm}\includegraphics[height=8cm]{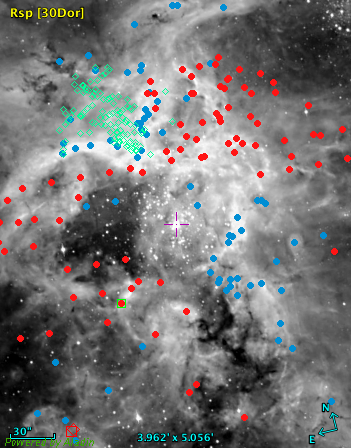}\vspace{0cm}\includegraphics[height=8cm]{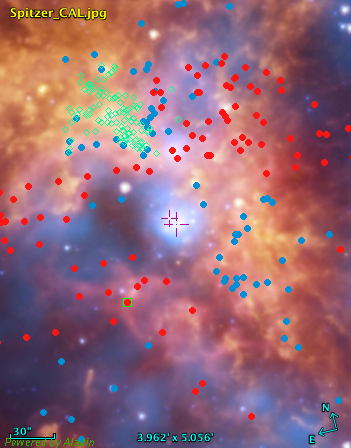}\includegraphics[height=8cm]{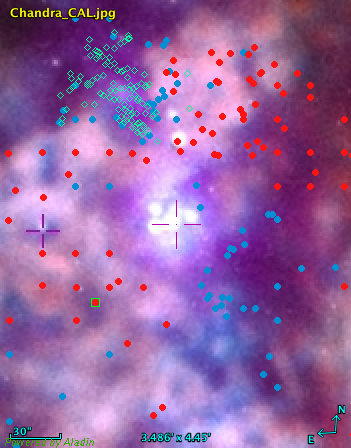}
\vspace{-0cm}
\caption {\small {\bf Left:} WFI \halfa image of 30Dor. {\bf Center:} Composite image from \cite{townsley}. The infrared emission from Spitzer is shown in inorange and the X-ray emission in blue. {\bf Right:} Chandra 20-year anniversary 3-channel X-ray image of the same region. The images have been astrometrically matched using point-like X-ray sources and the software package ALADIN. The Chandra image is rotated $\sim15^{\deg}$ clockwise relative to the others due to the ignorance of the lead author about the intricacies of ALADIN. As in Fig.\ref{fig110} in the main text, the red and blue dots show the positions of red and blue shifted single profiles; the green open rhombs mark positions of all the CO clumps identified by   \cite{inde13} in the giant molecular cloud 30Dor-10}
\label{app4}
\end{figure*}

 We discussed in the text (\ref{fig110}) how the patchiness of the X-ray emission is due to obscuration by nebular gas and dust, but we did not include an image of dust emission in that figure in order to avoid crowding the nebular features.  Figure~\ref{app4} shows the two views from Fig.\ref{fig110} plus a third panel showing  the dust emission from Spitzer. The GMC 30Dor-10 is shown in green, while as before, the red and blue dots represent blueshifted and redshifted nebular \halfa. The average radial velocity of 30Dor-10 is $<V_{H_2}>=-16$~\kms\footnote{There are two molecular clumps about 20pc west of the center of the cloud that have radial velocities close to V=0, similar to those of the single nebular lines in the same region.} indicating that it is either in front of R136, and being expelled from the nebular core by the radiation pressure, or behind and infalling towards R136.

The fact that 30Dor-10 is located exactly on top of a dark patch in the X-ray image suggests that the cloud is in front of. and obscures, the hot gas. However, the cloud also overlaps with one of the dustiest regions of 30Dor seen by Spitzer,  which is also quite luminous in \halfa.  The nebular gas in the region is also blue shifted, but the radial velocities do not match those of the GMC. Oposite to 30Dor-10, to the south of R136, there is a similarly luminous dust-emission and \halfa region, which however does not contain (significant) molecular gas. Therefore, it is reasonable to assume that most of the dust emission seen in the Spitzer image, including the region of 30Dor-10, that obscures the X-ray emission,. comes from nebular dust and not from dust associated with the GMC, and that 30Dor-10 is actually infalling towards R136.
 
If that is the case, the PDR models of  \cite{chevance}, which explicitly assume that dust is associated with the molecular cloud, are inaccurate, and the corresponding mass determinations are at best upper limits. The same is the case, of course, for the $H_2$ masses they derived from modeling the dust. As a corollary,  the discrepancy between the molecular mass from PDR modeling, and the CO masses of \cite{inde13}, is probably not as large as they claim. 

With the caveats above, \cite{chevance} estimate that the GMC stretches roughly radially between 20pc and 80pc from R136  so only fraction of the cloud would lie within the kinematical core ($R\leq25$pc). We will therefore assume that core contains at most half of the $H_2$ mass of 30Dor-10. i.e. $9\times10^4$\msun.

\end{document}